\PassOptionsToPackage{hyphens}{url}
\PassOptionsToPackage{obeyspaces}{url}
\PassOptionsToPackage{spaces}{url}

\documentclass[letterpaper,twocolumn,10pt]{article}
\usepackage{usenix-2020-09}

\let\OLDthebibliography\thebibliography
\renewcommand\thebibliography[1]{
  \OLDthebibliography{#1}
  \setlength{\parskip}{0pt}
  \setlength{\itemsep}{0pt plus 0.2ex}
}

\usepackage[utf8]{inputenc}
\usepackage{array, multirow}
\newcolumntype{P}[1]{>{\centering\arraybackslash}p{#1}}
\usepackage{tikz}
\usepackage{amsmath}
\usepackage{pifont}
\newcommand{\cmark}{\ding{51}} %
\newcommand{\xmark}{\ding{55}} %
\usepackage{makecell}
\usepackage{subcaption}
\usepackage{balance}
\usepackage{fancyhdr}
\usepackage[ruled,vlined]{algorithm2e}
\usepackage{authblk}

\expandafter\def\expandafter\UrlBreaks\expandafter{\UrlBreaks%
\do\a\do\b\do\c\do\d\do\e\do\f\do\g\do\h\do\i\do\j%
\do\k\do\l\do\m\do\n\do\o\do\p\do\q\do\r\do\s\do\t%
\do\u\do\v\do\w\do\x\do\y\do\z\do\A\do\B\do\C\do\D%
\do\E\do\F\do\G\do\H\do\I\do\J\do\K\do\L\do\M\do\N%
\do\O\do\P\do\Q\do\R\do\S\do\T\do\U\do\V\do\W\do\X%
\do\Y\do\Z\do\/\do-}

\usepackage{array}
\newcolumntype{H}{@{}>{\setbox0=\hbox\bgroup}c<{\egroup}}

\newcommand{\sad}{{SadDNS}}
\newcommand{\frag}{{FragDNS}}
\newcommand{\hijack}{{HijackDNS}}

\newcommand{\rfc}[1]{[RFC{#1}]}

\usepackage[labelfont=footnotesize,textfont=footnotesize]{caption}

\newcommand{\rot}[1]{\rotatebox[origin=c]{90}{ ~#1~ }}
\newcommand{\mr}[2]{ \multirow{#1}{*}{#2} }
\newcommand{\mcc}[2]{ \multicolumn{#1}{c|}{#2} }

\usepackage{tikz}
\newcommand*\circled[1]{\tikz[baseline=(char.base)]{
            \node[shape=circle,draw,inner sep=1pt] (char) {#1};}}
            
\usepackage[printwatermark]{xwatermark}
\newwatermark[allpages,color=black!100,angle=0,scale=1,xpos=0,ypos=-126]{\small 2021 30th USENIX Security Symposium \\ Accepted version. \url{https://www.usenix.org/conference/usenixsecurity21/presentation/dai}}

\begin{document}
\date{}
\title{\Large \bf The {\em Hijackers} Guide To The Galaxy: \\
  {\em Off-Path} Taking Over Internet Resources}

\author[*]{Tianxiang Dai}
\author[*$\dag$]{Philipp Jeitner}
\author[*]{Haya Shulman}
\author[*$\dag$]{Michael Waidner}
\affil[*]{Fraunhofer Institute for Secure Information Technology SIT}
\affil[$\dag$]{Technical University of Darmstadt}
\renewcommand\Authands{ and }

\maketitle

\begin{abstract}

Internet resources form the basic fabric of the digital society. They provide the fundamental platform for digital services and assets, e.g., for critical infrastructures, financial services, government. Whoever controls that fabric effectively controls the digital society.

In this work we demonstrate that the current practices of Internet resources management, of IP addresses, domains, certificates and virtual platforms are insecure. Over long periods of time adversaries can maintain control over Internet resources which they do not own and perform stealthy manipulations, leading to devastating attacks. We show that network adversaries can take over and manipulate at least 68\% of the assigned IPv4 address space as well as 31\% of the top Alexa domains. We demonstrate such attacks by hijacking the accounts associated with the digital resources. %

For hijacking the accounts we launch off-path DNS cache poisoning attacks, to redirect the password recovery link to the adversarial hosts. We then demonstrate that the adversaries can manipulate the resources associated with these accounts. We find all the tested providers vulnerable to our attacks.

We recommend mitigations for blocking the attacks that we present in this work. Nevertheless, the countermeasures cannot solve the fundamental problem - the management of the Internet resources should be revised to ensure that applying transactions cannot be done so easily and stealthily as is currently possible.

\end{abstract}

\section{Introduction}

Internet resources form the cornerstone of modern societies. The daily activities and services are increasingly digitalised, from critical infrastructures to medical services and child care. The society relies on the control over its Internet resources for availability and stability of digital services and assets. Due to their importance, Internet resources pose a lucrative target for adversaries.

{\bf Internet resources are at risk.} In this work we explore the security of the Internet management systems of basic digital assets: IP addresses management with Regional Internet Registries (RIRs) \rfc{7020}, domains with domain registrars, virtual machine resources with infrastructure as a service (IaaS) providers and certification with Certificate Authorities (CAs), see the list in Table \ref{password_recovery_table}. These providers manage the allocation, registration and operation of the Internet resources for their customers. We study how easy it is for network adversaries to take over the accounts of these resource providers and then exploit the resources associated with the compromised accounts.

We show that the current practices of Internet resources management are insecure. Adversaries can take control over digital assets of customers and maintain control over them for long periods of time without being detected. Although such attacks are appealing for strong nation state adversaries and security agencies, we demonstrate that even weak {\em off-path} network adversaries can, through a series of protocol manipulations, take over the accounts of customers and thereby control the Internet resources associated with them. 

{\bf Adversaries can hijack accounts.} The idea behind our attacks is the following: the adversary poisons the cache of the DNS resolver of a resource provider, injecting a malicious record mapping the Email server of the victim customer to an adversarial host. The adversary invokes password recovery procedure. The Email server of the provider sends the password reset link to the IP address of the adversary. Adversary resets the password and hijacks the victim account. We demonstrate how the adversary can perform manipulations over the resources associated with the hijacked accounts.

{\bf Manipulation of the digital resources.} The SSO (Single Sign On) accounts of the RIRs pose the highest risk: hijacking an SSO account allows a weak adversary to take over ASes and IP blocks allocated to the victim. Furthermore, through the hijacked account the adversary can make manipulations not only in the control plane of the Internet infrastructure but also in the Internet Routing Registries (IRR) and in Internet Addressing Resource Registries. Such modifications in the IRR can among others also facilitate extremely effective BGP prefix hijacks. Specifically, IRR records are prerequisite for BGP hijack attacks - without proper records in the IRR the attacker cannot convince benign upstream providers to accept and propagate the fraudulent BGP announcements in the input filters on the BGP sessions. Adversaries without the ability to modify the IRR, have to use less vigilant and generally poorly managed networks as upstream providers or have to utilise path manipulation attacks \cite{mitseva2018state} - both restricting the success rate and the stealthiness of the attack. Our adversary can, by modifying the records in the IRR, cause well managed and reputed upstream providers to unwittingly propagate the malicious BGP announcements. Hence making BGP prefix hijacks more effective than the typical control plane BGP prefix hijacks while at the same time more difficult to identify. To maintain control over the victim Local Internet Registries (LIRs) resources over long periods of time the adversary implants itself in the system with elevated privileges. 
 
We also show that hijacking an account under a CA allows an adversary to revoke certificates, renew a certificate or issue new certificates for domains registered under the hijacked account. Renewal of certificates allows to associate a new key-pair with the certificate. Nevertheless some CAs do not perform validation of certificate renewal requests issued from registered accounts. 

By hijacking the accounts of domain registrars, the adversary can manipulate records in victim domains, e.g., to launch phishing attacks. Finally, hijacking accounts of IaaS providers enables the attackers to take over virtual machines and the resources that run on those virtual machines, including databases, applications and computations.

{\bf Disclosure and ethics.} Our attacks were tested against providers and customers reliably, yet were ethically compliant. 
To avoid harming Internet users, we set up victim domains and registered victim accounts which were used by us for carrying out the attacks and for evaluating the vulnerabilities. This ensured that the providers would not use the spoofed records for any ``real'' purpose. In addition to evaluating the attacks with the ``victim'' accounts that we set up, we also evaluated our exploits of hijacked accounts against one large ISP under RIPE NCC and attacked the real domain of that ISP in coordination with that ISP. %
We are disclosing the results of our work to the providers.

{\bf Contributions.} We provide the first demonstration of off-path attacks exploiting hijacked accounts under popular providers and show that adversaries can perform manipulations in the resources assigned to the accounts over long time periods without being detected. %

{\bf Organisation.} %
In Section \ref{sc:cache:poison} we review DNS cache poisoning and related work. In Section \ref{sc:meth} we provide an overview of our study. In Section \ref{sc:dns:intercept} we list methodologies for off-path DNS cache poisoning attacks. In Section \ref{sc:rir} we evaluate the cache poisoning methodologies for taking over customers accounts in different providers in our dataset. Then, in Section \ref{sc:exploits}, we demonstrate how the adversaries can manipulate digital resources assigned to the accounts they control. In Section \ref{sc:digital:resources} we explain the fraction of the digital resources (IP address' blocks and domains) that are at immediate risk due to being associated with vulnerable accounts. We recommend countermeasures in Section \ref{sc:mitigations} and conclude in Section \ref{sc:conclusion}. %

\section{DNS Cache Poisoning Overview}\label{sc:cache:poison}

{\bf DNS.} Domain Name System (DNS), \rfc{1035}, performs lookup of services in the Internet. Recursive caching DNS resolvers receive DNS requests for services in different domains and send queries to the nameservers authoritative for those domains. The nameservers respond with the corresponding DNS records. The DNS records in responses are cached by the DNS resolvers and are provided to clients and servers which use that resolver. Subsequent requests for that domain are responded from the cache. For instance, to send an Email to \path{alice@example.info} the Email server of Bob will ask the DNS resolver for the IP address of the Email exchanger in domain \path{example.info}. The resolver asks the nameservers in domain \path{example.info} for an IP address and a hostname (\path{A} and \path{MX} records) of the Email exchanger and receives:

{\small
\begin{verbatim}
example.info IN MX mail.example.info
mail.example.info A 1.2.3.4
\end{verbatim}}

The resolver will send to the Email server of Bob the IP address of the Email exchanger of Alice and will also cache the records from the response for answering future queries for \path{MX} and \path{A} in \path{example.info} domain. 

{\bf DNS Cache Poisoning.} In a DNS cache poisoning attack the goal of the adversary is to redirect clients of some resolver to an IP address of the adversary for queries in a target victim domain. To do that, the adversary sends a DNS response from a spoofed source IP address, which belongs to the nameserver of the victim domain, with malicious DNS records mapping the victim domain to an IP address of the adversary. For instance, to intercept the Emails sent to Alice the adversary injects a DNS record mapping the Email exchanger of Alice to an adversarial host. If the resolver accepts and caches the malicious record, its cache becomes poisoned.

{\small
\begin{verbatim}
example.info IN MX mail.example.info
mail.example.info A 6.6.6.6
\end{verbatim}}

The added value of DNS cache poisoning attacks is that they have a local impact, affecting not the entire Internet but only the victim network and hence allow for extremely stealthy attacks, which can go undetected over long time periods. There is more and more evidence of DNS cache poisoning in the wild and the attacks are becoming increasingly sophisticated. In the recent cache poisoning attacks in the wild the adversaries attempt to intercept DNS packets by launching short-lived BGP (Border Gateway Protocol) prefix hijacks \cite{bgp:attacks}. In such attacks, the adversary advertises a BGP announcement hijacking the prefix of a victim for a short time only to hijack the target DNS packet and then releases the hijack \cite{demchak2018china}. This allows the attacker to poison the DNS cache of a victim resolver and then intercept all the communication between the victim resolver and the target domain. Recent research projects showed that the CAs (Certificate Authorities) and the bitcoin infrastructures were not resilient to prefix hijacks \cite{birge2018bamboozling,birgeexperiences,apostolaki2017hijacking}. 

{\bf History of DNS Cache Poisoning.} Launching cache poisoning in practice is however hard. We explain the evolution of cache poisoning attacks and the mitigations. In 1995 Vixie pointed out to the cache poisoning vulnerability and suggested to randomise the UDP source ports in DNS requests \cite{Vixie95}. In 2002 Bernstein also warned that relying on randomising Transaction ID (TXID) alone is vulnerable \cite{Bernstein:DNS}. Indeed, in 2007 \cite{klein2007bind} identified a vulnerability in \textsf{Bind9} and in \textsf{Windows DNS} resolvers \cite{klein2007windows} allowing off-path attackers to reduce the entropy introduced by the TXID randomisation. In 2008 Kaminsky \cite{Kaminsky08} presented a practical cache poisoning attack even against truly randomised TXID. Following Kaminsky attack DNS resolvers were patched against cache poisoning \rfc{5452} by randomising the UDP source ports in queries. Nevertheless, shortly after different approaches were developed to bypass the source port and the TXID randomisation for launching off-path cache poisoning attacks. In 2012 \cite{conf/esorics/HerzbergS12} showed that off-path adversaries can use side-channels to infer the source ports in DNS requests. In 2015 \cite{shulman2015towards} showed how to attack resolvers behind upstream forwarders. This work was subsequently extended by \cite{zheng2020poison} with poisoning the forwarding devices. A followup work demonstrated such cache poisoning attacks also against stub resolvers \cite{collaborative-poison}. \cite{saddns} showed how to use ICMP errors to infer the UDP source ports selected by DNS resolvers. Recently \cite{klein2020cross} showed how to use side channels to predict the ports due to vulnerable PRNG in Linux kernel. In 2013 \cite{cns:frag:dns} provided the first feasibility result for launching cache poisoning by exploiting IPv4 fragmentation. IPv4 fragmentation based attacks were applied to shift time on NTP servers \cite{malhotra_attacking_2016,brandt2018domain,ntp-over-dns}, these attacks are not practical anymore since the nameservers in NTP domains were patched to avoid fragmentation. The study in \cite{brandt2018domain} used fragmentation based cache poisoning for bypassing domain validation with CAs. However, most CAs patched the vulnerabilities which \cite{brandt2018domain} exploited to attack domain validation, e.g., Let'sEncrypt blocked fragmentation. Let'sEncrypt also deployed domain validation from multiple vantage points \cite{multi:va,birgeexperiences}, which makes the previous off-path attacks \cite{brandt2018domain,birge2018bamboozling} impractical. 

In addition to other attacks in this work, we also show another way to attack the CAs, by taking over customers' accounts with the CAs and not by bypassing domain validation. As we show this allows even more effective attacks that were presented in \cite{brandt2018domain}: (1) when controlling a compromised account the adversary can renew existing certificates to use a new key-pair. Since some CAs do not apply domain validation during certificates' renewal this attack allows to issue fraudulent certificates without the need to attack DV. Furthermore, in our work we use a number of off-path DNS cache methodologies from \cite{dai:usenix:2021} to take over accounts with providers.

Cache poisoning attacks could be prevented with DNSSEC [RFC6840] \cite{weiler2013rfc} which uses cryptographic signatures to authenticate the records in DNS responses. However, DNSSEC is not widely deployed. Less than 1\% of the second level domains (e.g., 1M-top Alexa) domains are signed, and most resolvers do not validate DNSSEC signatures, e.g., \cite{chung2017longitudinal} found only 12\% in 2017. Our measurements show that the DNSSEC deployment in our datasets is not better: the resolvers of 19 out of 35 tested providers do not validate DNSSEC signatures (see Table \ref{resolvers_accepting_fragments_table}) and less than 5\% of the customers' domains are signed. Deploying DNSSEC was showen to be cumbersome and error-prone \cite{chung2017understanding}. Even when widely deployed DNSSEC may not always provide security: a few research projects identified vulnerabilities and misconfigurations in DNSSEC deployments in popular registrars \cite{shulman2017one,chung2017longitudinal}. However, even correctly deployed DNSSEC does not prevent recent cache poisoning attacks \cite{272204}. The idea behind these attacks is to encode injections into the DNS records in DNS responses. When the resolver parses the records, a misinterpretation occurs, such that when the record is stored a different domain name is used. Since DNSSEC validation is applied prior to the misinterpretation, the validation of DNSSEC succeeds, and the DNS cache poisoning occurs afterwards. Preventing these attacks requires validating or escaping records from DNS lookups.

Recent proposals for encryption of DNS traffic, such as DNS over HTTPS \cite{hoffman2018rfc} and DNS over TLS \cite{hu2016rfc}, although vulnerable to traffic analysis \cite{shulman2014pretty,siby2019encrypted}, may also enhance resilience to cache poisoning. These mechanisms are not yet in use by the nameservers in the domains that we tested. Nevertheless, even if they become adopted, they will not protect the entire resolution path, but only one link on which the transport is protected and hence will not completely prevent DNS cache poisoning attacks.

\section{Attack Overview}\label{sc:meth}

In our study we explore the security of the services which provide access to and management of the key digital assets in the Internet: domains, IP prefixes and ASes, virtual machines and certificates. In Table \ref{password_recovery_table} we list the resources, as well as the public service providers of these resources, that we studied in this work. Access and management of these digital resources is performed with the accounts that the providers offer to the customers via their web portals. In this section we provide an overview of our study from the perspective of the adversary for hijacking the accounts of the customers under different resource providers.  %

{\bf Find the target.} Assume for example that the adversary wishes to hijack the DNS servers hosted on a victim prefix {\tt 205.251.192.0/18} -- this was a real attack launched against an LIR Amazon route53 in April 2018. First, the adversary needs to find an account to which these resources are assigned and through which these resources can be managed. Then, the adversary needs to find the username associated with that account. In Section \ref{sc:passwords} we show how to find the needed information: the owner, the public service provider, the Email which is associated with the account through which the digital resources can be managed. In the case of our example, the prefix is allocated by ARIN to an LIR with OrgId \texttt{AMAZON-4}, aka Amazon.com, Inc. and has 3 origin ASNs (Autonomous System Numbers) registered: AS16509, AS39111 and AS7224. We thereby learn that the responsible RIR for Amazon is ARIN and that Amazon has an LIR agreement with ARIN. We also find the Email address \path{ipmanagement@amazon.com} used by Amazon for managing its resources via the SSO account with ARIN.

\begin{figure}[ht!]
\vspace{-5pt}
    \centering
    \includegraphics[width=0.43\textwidth]{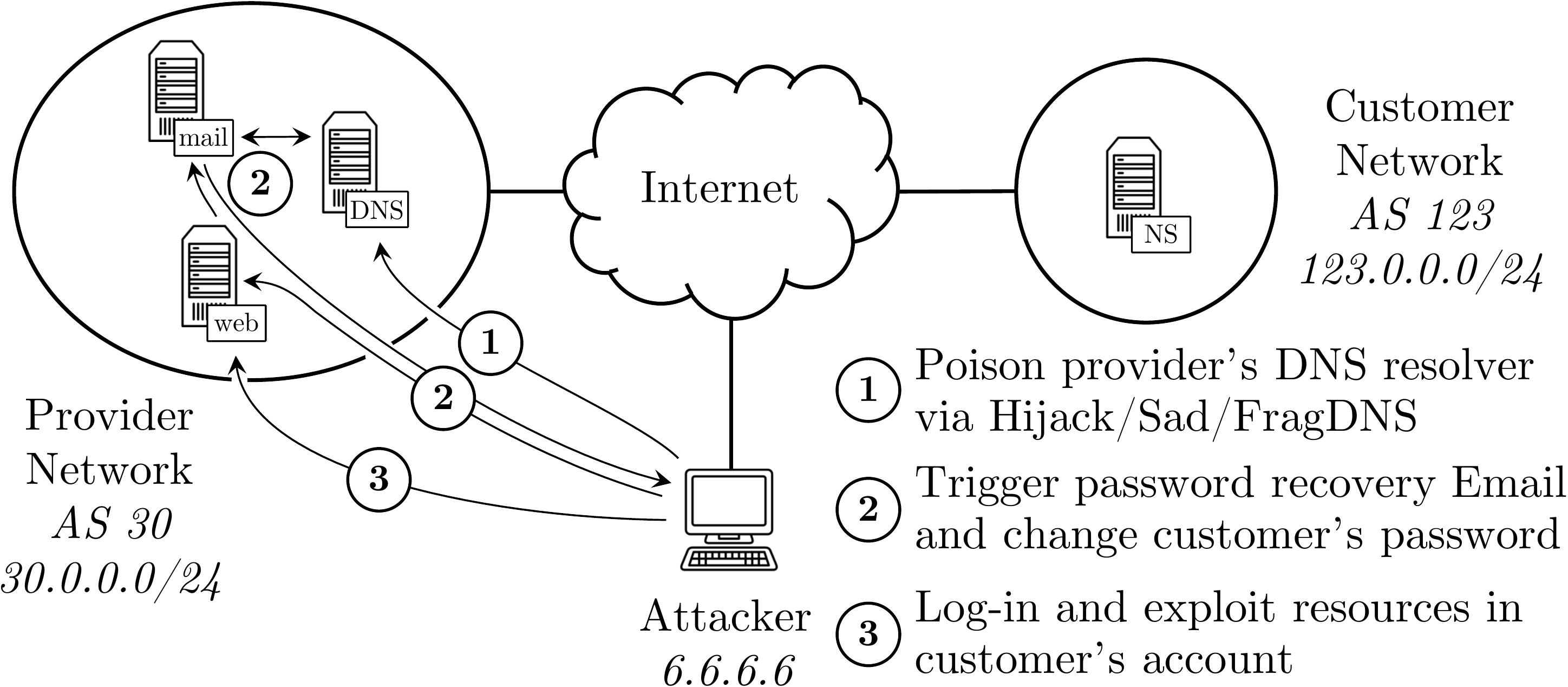}
    \caption{Attack overview}
    \label{fig:sequence_overview}
\vspace{-10pt}
\end{figure}

{\bf Poison DNS of public service provider.} The adversary uses one of the methodologies in Section \ref{sc:dns:intercept} to launch off-path DNS cache poisoning attack against the DNS resolver of the service provider ARIN. During the attack the adversary injects a malicious DNS record mapping the Email server of domain \path{amazon.com} to the IP addresses controlled by the adversary (step \circled{1}, Figure \ref{fig:sequence_overview}). As a result, the Emails sent by ARIN to Amazon will be received by the adversary. %

{\bf Hijack victim account.} The adversary triggers password recovery for Email \path{ipmanagement@amazon.com}. This Email is associated with the SSO account at ARIN. In order to send the password recovery link, the Email server at ARIN needs the IP address of the Email server of Amazon. The resolver at ARIN already has a corresponding record in the cache, which it provides to the Email server. This IP address was injected by the adversary earlier in step \circled{1}. ARIN sends the Email with password recovery instructions to the adversary (step \circled{2}, Figure \ref{fig:sequence_overview}). The attacker resets the password and takes control over the account. We experimentally evaluate such attacks against the providers and their customers in our dataset in Section \ref{sc:rir} for details. 

{\bf Manipulate the resources.} The adversary manipulates the resources assigned to the victim account, say of Amazon, and can sell the IP prefixes and ASes owned by Amazon (step \circled{3}, Figure \ref{fig:sequence_overview}). In Section \ref{sc:exploits} we describe the exploits we evaluated against the resources assigned to our victim accounts. We show that among others, the attacker can create additional accounts for itself with arbitrary privileges, and hence even if the real owner resets the password back, the attacker still maintains control over the resources. In some cases these manipulations generate notification Emails to the Email address associated with the resources. This Email address is however hijacked by the adversary, hence the adversary receives the notifications. As a result the attack will not be detected and can stay under the radar over a long period of time. %

\section{Off-Path DNS Cache Poisoning}\label{sc:dns:intercept}
The key contribution in our work is to show that once an adversary controls an account with a resource provider, it can in an easy and stealthy way manipulate the digital resources associated with that account. But, how easy is it to take over accounts? We show how to take over accounts by injecting a poisoned DNS record into the caches in DNS resolvers of providers. When the adversary triggers the password recovery procedure for the victim account, the reset email is sent to the adversarial host at the IP address in the injected DNS record. 

How easy is it to launch off-path DNS cache poisoning? In this section we use methodologies from \cite{dai:usenix:2021} to launch off-path DNS cache poisoning attacks: BGP prefix hijacks \cite{birge2018bamboozling}, side-channels \cite{saddns}, and IPv4 defragmentation cache poisoning \cite{cns:frag:dns}. We do not consider attack methodologies which are effective only against specific operating systems, say due to poor random number generators. We implement cache poisoning attacks using these methodologies and evaluate them against the providers and the customers in our dataset. We describe the experimental setup in Section \ref{sc:setup}. We explain our study methodology in Section \ref{sc:study:methodology}. Then in Sections \ref{sc:hijack}, \ref{sc:saddns} and \ref{sc:fragDNS} we present the DNS cache poisoning methodologies and the experimental evaluations against the targets in our dataset.

\subsection{Setup}\label{sc:setup}
To test our attacks experimentally in the Internet we setup a victim AS. To purchase the victim AS we registered a secondary LIR account with RIPE NCC for our organisation (which has a primary account with RIPE NCC). We purchased a /22 prefix for our AS for 20,000USD. We connected our AS to DE-CIX internet exchange point in Frankfurt. This AS hosts the servers which we use for our evaluation of the attacks. 

We set up an \textsf{Unbound} 1.6.7 DNS resolver on Linux 4.14.11, whose cache we poison with the records of the customer domains. We registered a victim domain and set up two nameservers in our domain and an Email server. We use our victim domain to register accounts with the services that we test in this work. We call this domain the victim customer domain. We also set up a border router which represents our attacker. The attacker's BGP router issues bogus BGP announcements that claim the prefix assigned to our victim AS. This allows us to evaluate the viability of attacks with BGP prefix hijacks against our domains hosted on our victim AS without affecting services and domains not under our control and without affecting the global BGP routing table in the Internet. 

To evaluate cache poisoning attacks with side-channels we configure the nameservers in our domain to support rate-limiting and the DNS resolver to issue ICMP errors. To evaluate fragmentation based cache poisoning attacks we configure nameservers in our domain to reduce the MTU according to the value in ICMP fragmentation needed messages. The nameservers in our victim domain use a globally incremental IPID counter.

\subsection{Study Methodology}\label{sc:study:methodology}
Our experimental evaluation of the attacks is performed reliably yet without disrupting the functionality of the customers. To achieve this we evaluate the attacks in two steps: (1) We evaluate vulnerabilities to cache poisoning in providers. For this we set up victim domains and register victim accounts with the providers. We experimentally test the attack methodologies against providers by poisoning their DNS caches with malicious records mapping the Email server in our victim domain to the adversarial hosts that we control. We then hijack our victim accounts by triggering the password recovery procedures and changing their passwords. This enables us to validate vulnerabilities to cache poisoning yet without risking that the providers use poisoned records for genuine customers. The ability to take over the accounts of the real customers depends not only on vulnerabilities in providers' infrastructure but also on properties in customers' domains. (2) Hence, in this step we set up a victim DNS resolver and poison its cache with malicious records mapping the genuine customer domains to our adversarial hosts. The combination of both evaluations against the providers and against the customers enables us to estimate the extent of the vulnerable accounts that can be hijacked.

\subsection{BGP Prefix Hijack}\label{sc:hijack}\label{sc:bgp:eval}
BGP (Border Gateway Protocol) allows ASes to compute the paths to any Internet destination. %
Since BGP is currently not protected, adversaries can send bogus BGP announcements to hijack victim prefixes, hence intercept the communication of victim ASes that accept the malicious BGP announcements. In our attacks we hijack the prefix of our AS: once in the evaluation against providers to intercept the responses from our nameservers sent to the DNS resolvers of the providers and then again during the evaluation of the customers, to intercept requests from our DNS resolver to the customers' domains. After our AS accepts the bogus BGP announcement, all the communication between the servers on our AS and the servers of the targets in our dataset traverse our adversarial BGP router.

We launch short-lived hijacks. Such hijacks are common \cite{rexford2002bgp} and allow the attacker to stay below the radar \cite{crypto:currency,dns:venezuela}. It is believed that short-lived traffic shifts are caused by the configuration errors (that are quickly caught and fixed) and since they do not have impact on network load or connectivity, they are largely ignored \cite{boothe2006short,karlin2008autonomous,khare2012concurrent}. We evaluate our attacks using short-lived same prefix hijacks and sub-prefix of the victim prefix.

Our experimental evaluation reflects a common BGP hijacking attacker: the attacker controls a BGP router or an AS, and issues BGP announcements hijacking the same-prefix or a sub-prefix of a victim AS in the Internet. 

\subsubsection{Attack evaluation against providers}
The adversary announces to our victim AS a prefix of the network of the provider where the target DNS resolver is located. The bogus BGP announcement is sent {\em only} on the interface that is connected to our AS and is not sent to other destinations in the Internet. As a result, the responses from the nameservers of our victim domain are sent to the adversarial host instead of the DNS resolver of the provider. The adversary initiates password recovery procedure for an account of our victim customer domain. This triggers a DNS request to our victim domain. The corresponding nameserver sends a response, which is instead redirected to the adversary's host. The adversary manipulates the response, and injects a DNS record that maps the Email server of our victim domain to the IP address of the adversary. The response is then sent to the provider and the BGP hijack is released. The DNS resolver caches the response and returns it to the Email server, which sends the password recovery link to the IP address of our adversary. The adversary resets the password and takes control over the account.

\subsubsection{Attack evaluation against customers}
The adversary announces to our victim AS prefixes of the networks that host the nameservers in the target customers' domain. The bogus BGP announcements are sent {\em only} on the interface that is connected to our AS and not to other destinations in the Internet. As a result, the DNS requests from the DNS resolver on our victim AS are sent to the adversarial host instead of the nameservers of the customer' domain. The attacker releases the hijacked prefix, and additionally crafts a spoofed DNS response to our DNS resolver mapping the IP address of the adversary to the Email server of the victim customer's domain. The records from the DNS response are cached by our resolver.

\subsection{Side-channel Port Inference}\label{sc:saddns}\label{sc:sad:eval}

SadDNS off-path attack \cite{saddns} uses an ICMP side channel to guess the UDP source port used by the victim resolver in the query to the target nameserver. This reduces the entropy in a DNS request from 32 bit (DNS TXID \& UDP port) to 16 bit. The adversary then uses brute-force to match the TXID by sending spoofed packets for each possible TXID value to the resolver.

\subsubsection{Attack evaluation against providers}
We verify the existence of the ICMP global-rate limit: we send a single UDP probe to the resolver to verify that it emits ICMP port unreachable messages. Then, we send a burst of 50 spoofed UDP packets to closed ports at the resolver and follow up with a single non-spoofed UDP packet and observe if an ICMP port unreachable message is received by our sender. If the ICMP global rate-limit is present no message will be received because the global rate-limit is already reached.

The adversary initiates password recovery procedure with a provider for an account of our victim customer domain. This triggers a DNS request to our victim domain. The adversary mutes the nameservers on our victim AS, to prevent the response from being sent to the resolver of the provider, then runs the procedure for inferring the source port in the DNS request. Once the source port is found, it sends $2^{16}$ spoofed responses for each possible TXID values with malicious DNS records in payload. The records map the nameservers of the victim domain to attacker controlled IP addresses. If the response is accepted by the resolver of the public service, it is cached and used by the service for sending an Email with the password or the reset link. The attacker now controls the account.

\subsubsection{Attack evaluation against customers}%
We configure our DNS resolver to send ICMP errors on closed ports. We use our own implementation of the \sad\ port scanning application with binary search and attempt to poison the resolver with a malicious record pointing the domain of the customer to our adversarial host. Due to a high failure rate, evaluation of each tuple (resolver, domain) takes up to 30 minutes, hence evaluating \sad\ on all the domains in our dataset is not practical. We therefore perform the measurement on a dozen randomly selected customers in our dataset. Our implementation performs the complete attack from triggering the queries to muting the nameservers and scanning the ports (using the ICMP side-channel) and in the last step sending the spoofed DNS responses with malicious records.

The high failure rate of the \sad\ attack is due to the fact that most of the queries do not generate a useful attack window, since the resolver times out after less than a second. The attacker can further improve this via manual attack by analysing the back-off strategies of the target resolver. The timeout of the resolver is implementation dependent, e.g., the timeout value of \textsf{Unbound} is a dynamically computed value based on RTT to the nameserver, while \textsf{Bind} uses 0.8 seconds. The DNS software increases the timeout value after each retransmission.

\subsection{Injection into IP-Defragmentation Cache}\label{sc:fragDNS}\label{sc:frag:eval}
The off-path adversary uses a spoofed IPv4 fragment to manipulate the fragmented response from the nameserver, \cite{cns:frag:dns}. The idea is to send a spoofed fragment which is reassembled with the first genuine fragment of the nameserver. The adversary replaces the second fragment of the nameserver with its malicious fragment, hence overwriting some parts of payload of a DNS response with new (attacker's injected) content. As a result, the reassembled IP packet contains the legitimate DNS records sent by the genuine nameserver with the malicious records from the fragment sent by the adversary. Since the challenge values (port, TXID) are in the first fragment of the response from the nameserver, they remain intact, and hence correct.

\subsubsection{Attack evaluation against providers}
We evaluate \frag\ attack against the resolvers of the providers with our victim domain. For our nameservers we use a custom application that we developed, which always emits fragmented responses padded to a certain size to reach the tested fragment size limit. The nameservers are configured to send \path{CNAME} records in the first fragmented response. As a result, when the resolver of the provider receives a fragmented response and reassembles it, the DNS software will issue a subsequent query for the \path{CNAME}-alias. This allows us to verify that the spoofed fragment arrived at the resolver and was reassembled correctly and cached, which is an indicator that the cache poisoning via fragmentation attack succeeded. Throughout the attack we use our adversarial host to trigger password recovery procedures and to inject malicious DNS records into the caches of the providers, mapping the Email servers in our domains to the IP addresses allocated to our adversary.

The adversary sends two spoofed fragments (for each nameserver's IP address) to the resolver of the public service. The fragments are identical except for the source IP addresses: one is sent with a source IP address of one nameserver and the other is with the source IP address of the other nameserver. The fragments are constructed so that they match the first fragment in the response that will have been sent by our nameserver. In the payload the fragments contain malicious DNS records mapping the Email server to the IP address of the adversary. The adversary initiates password recovery for our victim account. This triggers a DNS request to one of the nameservers in our victim domain. We do not know in advance which nameserver that will be, and hence initially send two spoofed fragments (for each nameserver). The response from the nameserver is sent in two fragments. Once the first fragment reaches the IP layer at the resolver of the provider it is reassembled with one of the second fragments of the adversary (it is already waiting in IP defragmentation cache). The reassembled packet is checked for UDP checksum and if valid, passed on to the DNS software on the application layer. If the records from the DNS packet are cached by the resolver of the public service, the password recovery link will be sent to the host controlled by the attacker.

\begin{figure}[ht]
	\centering
	\includegraphics[width=0.48\textwidth]{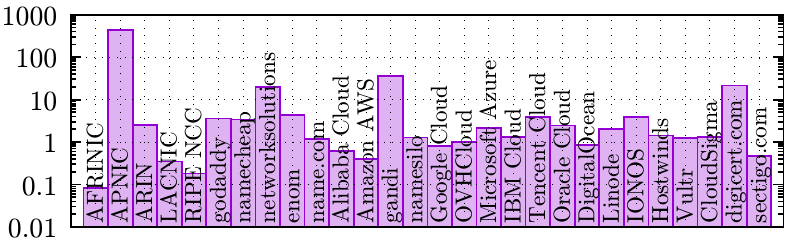}
	\vspace{-20pt}
	\caption{Avg. latency (in seconds) between registration and resolver query, excluding outliers outside $\pm{}1\sigma$.}
	\label{fig:latency}
	\vspace{-5pt}
\end{figure}

{\bf When to send the spoofed `second' fragment?} The standard [RFC791] recommends caching the IP fragments in IP defragmentation cache for 15 seconds. If there is no matching fragment after 15 seconds, the fragment is removed from IP defragmentation cache. The actual caching time exceeds 15 seconds in most implementations. For instance in Linux \path{/proc/sys/net/ipv4/ipfrag_time} is set to 30 seconds. For attack to succeed the total time between the moment that the fragment enters the IP defragmentation cache at the provider's resolver and the moment at which the first fragment from the genuine nameserver arrives should not exceed 15 seconds (to ensure that the attack is effective not only against resolvers running on Linux but also against standard compliant operating systems). We show that in practice 15 seconds suffice to launch the attack. We measure the latency from the moment that we trigger the password recovery procedure via the web interface of the provider and the moment that a DNS request from the resolver of the provider arrives at our nameservers. The measurements for different providers are plotted in Figure \ref{fig:latency}. As can be seen, except for two providers, all the latencies are below 30 seconds. 
The results for the attack window across all the providers, plotted in Figure \ref{fig:latency} show that the latencies are stable, and are within the interval which provides for successful attacks. For instance, for AFRINIC RIR the attacker learns that after issuing the password recovery procedure the DNS query will be sent to the victim nameserver at a predictable time interval (between 0.1 and 0.2 seconds).

\subsubsection{Attack evaluation against customers}
Our evaluation is performed with the domain of a victim customer against our DNS resolver. The DNS resolver is configured to allow fragmentation. We look-up the nameservers in the domain of the customer and check if we can force them to fragment responses: (1) for each nameserver, our DNS resolver sends requests to the nameserver and receives responses. (2) From the adversarial host we send to these nameservers {\em ICMP fragmentation needed} errors indicating Packet Too Big (PTB) for the source IP address of our DNS resolver. (3) We send DNS requests from our resolver and check if the responses arrive fragmented according to the MTU indicated in the ICMP errors. 

We then run \frag\ attack against the nameservers that fragment DNS responses following our ICMP PTB errors: (4) The adversarial host crafts spoofed second fragments, one for each nameserver in customer's domain. Since the adversary does not know to which nameserver the resolver will send a DNS request (the nameserver selection depends on DNS resolver software) it will send spoofed second fragments for each of the nameserver in that domain. Each fragment contains an identical payload: a malicious DNS record that maps the Email server of the customer domain to the IP address of our adversary. Each fragment has a different spoofed source IP address corresponding to each of the nameserver in the target domain. The adversary sends all these fragments to our DNS resolver. (5) The adversary causes our DNS resolver to issue a DNS request for a \path{MX} record in victim customer's domain. The nameserver which received the request responds with a fragmented DNS packet. The first fragment is reassembled with the matching second fragment that is waiting in the IP defragmentation cache. (6) The adversary receives a DNS response from our resolver. If the Email server in the response is mapped to the IP address of our adversary, then the attack succeeded.

 \section{Hijacking Accounts}\label{sc:rir}

In this section we evaluate DNS poisoning attacks against the providers and the customers using the methodologies in Section \ref{sc:dns:intercept}. %

After collecting the target providers and their customers in Section \ref{sc:dataset}, we analyse the password recovery mechanism at each provider in Section \ref{sc:passwords}. Then we collect the DNS resolvers at those providers in Section \ref{sc:vic:rir}. We evaluate off-path cache poisoning attacks against the DNS resolvers of providers in Section \ref{ssc:eval:poson}. Finally we measure the percentage of vulnerable customers of those providers in Section~\ref{sc:vuln-customers}.

\subsection{Datasets}
\label{sc:dataset}
In our measurements and attacks' evaluations we use two datasets: of providers and of their customers.

{\bf Providers.} The providers that we study are listed in Table~\ref{password_recovery_table}. For each class of resource providers (RIRs, Registrars, IaaS providers, CAs) we select a set of most popular examples. Our methodology for selecting the providers is: (1) all the five RIRs, (2) we scan the {\tt whois} data of 100K-top Alexa domains and select the top 15 registrars according to the number of domains each registrar is managing, (3) to select the IaaS providers, we use market share data and supplement it with additional selected providers\footnote{\url{https://www.srgresearch.com/articles/quarterly-cloud-spending-blows-past-30b-incremental-growth-continues-rise}, \url{https://stackify.com/top-iaas-providers/}, \url{https://www.g2.com/categories/infrastructure-as-a-service-iaas}}, (5) we select the top 5 CAs which cover 97\% of the market share\footnote{\url{https://w3techs.com/technologies/history_overview/ssl_certificate}: this market share data lists most of Let'sEncrypt certificates as issued by IdenTrust as Let'sEncrypt certificates are cross signed by IdenTrust. We do not test Let'sEncrypt itself because it does not offer traditional user accounts and therefore does not support password recovery.}, all other CAs have less than 1\% market share. 

For registrars and IaaS providers these datasets include providers which we could not test, because they do not allow creation of user accounts. For example, publicdomainregistry does not offer accounts to end-users directly, but only manages domain registration for webhosters. Providers where we could not register accounts are: tucows.com, publicdomainregistry, cscglobal, markmonitor, Rackspace cloud, CenturyLink Cloud and Joyent Triton.

We obtain a list of 32 resource providers which use 1,006 resolvers for sending Email (back-end IP addresses) on 44 ASes associated with 130 prefixes. Some resource providers use only a small amount of Email servers and resolvers on their own networks, while other providers use large pools of Email servers and resolvers provided by third-party Email services like Mailchimp and Sendgrid. We list this technical information in Table~\ref{resolvers_accepting_fragments_table}.

{\bf Customers.} We extract account information for customers of RIRs and domain registrars from {\tt whois} databases. We parse the Email addresses in {\tt whois} records to extract the domains of the customers and query the nameservers responsible for those domains. 

Because of data protection settings, not all {\tt whois} records contain Email addresses, or only contain masked Email addresses which point to a registrar's Email proxy. We were able to find Email addresses for 74.62\% of the ASes from RIR {\tt whois} databases and for 10.60\% of the domains owners in 100K-top Alexa list from domain registrar {\tt whois} databases (see Table~\ref{rir_whois_database_details}). We collected 94,997 user accounts hosted in 59,322 domains and 69,935 nameservers. 

We were not able to retrieve user account information for IaaS accounts and CAs as this is not possible ethically in an automated way. An adversary can obtain this information, e.g., by enumerating usernames as described in Section~\ref{sc:passwords}. 

Our dataset of domain registrars is also representative for other types of resources hosted under that domain. Organisations which own domains also own cloud resources at IaaS providers and certificates at CAs and use the same domain for their Email addresses and therefore are vulnerable to the same attack at those providers.

\subsection{Collecting Accounts' Information} %
\label{sc:passwords}

The first step in our attack is to trigger the password recovery procedure at the provider. This step requires collecting information of the target customer whose account the attacker attempts to hijack, such as the Email account required to log into the target account, a username or a handle. We study for each service provider which information is needed for password recovery and how to collect that information for our targets; the data is summarised in Table \ref{password_recovery_table}.
We found that the customers' Email addresses can often be retrieved from the public {\tt whois} records. We were able to extract the Email addresses associated with the accounts at the providers for 41\% of the customers in our study. For instance, the Emails for the SSO accounts of 74.62\% of the LIRs (i.e., the customers of RIRs) can be retrieved via {\tt whois}.

For victim customers whose details cannot be publicly accessible via {\tt whois} we find the required information with manual research and dictionary attack. To carry out the dictionary attack we used the observations we derived from our data collection from public sources: the data we collected through {\tt whois} shows that more than 24\% of the Email addresses use one of ten well-known username parts, like \texttt{domains@email.info},  \texttt{hostmaster@email.info}, etc., which enables an informed attacker to find the Email addresses in less than ten attempts when these details are not publicly available through {\tt whois}. We apply dictionary attack to also recover other details: for example, our study shows that about 1 in 10 LIRs (customers of RIRs) use usernames that are identical to the Email address that is registered in the {\tt whois} records; e.g., username \texttt{operator} is associated with Email address \texttt{operator@email.info}.

\begin{table}[!ht]

\newcommand{\nousr}{$^2$}

\begin{center}
\footnotesize
\setlength\tabcolsep{2.5pt}

\begin{tabular}{|r|l|ccc|cccHH|}
    \hline
    \rot{Type} & Provider               & \makecell{Details needed\\for PW recovery} & \rot{Public-known}   & \rot{ ~Captcha~ } & \rot{Fragment} & \rot{SadDNS} & \rot{BGP hijack} & \rot{Override} & \rot{Mail-Spoofing} \\
    \hline
    \multirow{5}{*}{\rot{RIRs}}
         & AFRINIC                & NIC-handle                             & \cmark & \xmark & \cmark & \xmark & \cmark & \cmark & \cmark \\ %
         & APNIC                  & Email                                 & \cmark & \xmark & \cmark & \xmark & \xmark & \cmark & \cmark \\ %
         & ARIN                   & Email, Username                       & \xmark & \cmark & \cmark & -      & \xmark & \cmark & \cmark \\ %
         & LACNIC                 & Username                               & \cmark & \xmark & \cmark & \xmark & \cmark & \cmark & \cmark \\ %
         & RIPE NCC               & Email                                 & \cmark & \cmark & \cmark & \xmark & \cmark & \cmark & \cmark \\ %
    \hline
    \multirow{11}{*}{\rot{Domain registrars}}
         & godaddy                & Email, Domain name                    & \cmark & \xmark & \cmark & -      & \cmark & \cmark & \cmark \\ 
         & namecheap              &    Email                              & \cmark & \cmark & \cmark & \xmark & \cmark & -      & \xmark \\
         & networksolutions       &  Email                                & \cmark & \xmark & \cmark & \xmark & \cmark & -      & \cmark \\ 
         & enom.com               & Login ID, Sec. question                & \xmark & \cmark & \cmark & \xmark & \cmark & -      & \cmark \\
         & name.com               & Username${}^1$                         & \cmark & \xmark & \cmark & -      & \cmark & -      & \cmark \\
         & Alibaba Cloud          & Username, 2-FA                         & \xmark & \cmark & \cmark & \xmark & \cmark & -      & \cmark \\
         & Amazon AWS             &  Email                                & \cmark & \cmark & \cmark & \xmark & \cmark & -      & \xmark \\
         & gandi.net              &  Email                                & \xmark & \xmark & \cmark & \xmark & \cmark & \cmark & \cmark \\
         & namesilo.com           &  Email, Sec. question                 & \xmark & \cmark & \cmark & \xmark & \cmark & \cmark & \xmark \\
         & Google Cloud           & Last password, 2-FA                    & \xmark & \cmark & \cmark & \xmark & \cmark & -      & \xmark \\ %
         & ovh.com                & Email                                 & \cmark & \xmark & \cmark & \xmark & \cmark & -      & \cmark \\ 
    \hline
    \multirow{14}{*}{\rot{Cloud management (IaaS)}}
         & Amazon AWS             & Email                         & \multirow{14}{*}{\rot{Typically no for all IaaS providers}}
                                                                                    & \cmark  & \cmark & \xmark & \cmark & -      & \xmark \\ %
         & Microsoft Azure        & Email                                 &        & \xmark  & \xmark & \xmark & \cmark & -      & \xmark \\ 
         & Alibaba Cloud          & Username, 2-FA                         &        & \xmark  & \cmark & \xmark & \cmark & -      & \cmark \\ %
         & Google Cloud           & Last password, 2-FA                    &        & \xmark  & \cmark & \xmark & \cmark & -      & \xmark \\ %
         & IBM Cloud              & Email ('id')                          &        & \xmark  & \cmark & \cmark & \cmark & \xmark & \cmark \\ 
         & Tencent Cloud          & Email                                 &        & \cmark  & \cmark & \xmark & \cmark & -      & \cmark \\ 
         & Oracle Cloud           & Email                                 &        & \xmark  & \xmark & \xmark & \cmark & \cmark & \cmark \\ 
         & DigitalOcean           & Email                                 &        & \xmark  & \cmark & \xmark & \cmark & \xmark & \xmark \\ 
         & Linode                 & Username${}^1$                         &        & \xmark  & \cmark & \cmark & \cmark & \xmark & \xmark \\ 
         & IONOS                  & Email, id or domain                   &        & \xmark  & \cmark & \cmark & \cmark & -      & \cmark \\ 
         & Hostwinds              & Email                                 &        & \xmark  & \xmark & \xmark & \cmark & \cmark & \xmark \\ 
         & OVHcloud               & Email                                 &        & \cmark  & \cmark & \xmark & \cmark & -      & \cmark \\ %
         & Vultr                  & Email                                 &        & \cmark  & \cmark & \cmark & \cmark & -      & \xmark \\ 
         & CloudSigma             & Email                                 &        & \xmark  & \cmark & -      & \cmark & \xmark & \xmark \\ 
    \hline
    \multirow{5}{*}{\rot{CAs}}
    & IdenTrust  & Account number                                          & \xmark & \xmark  & \cmark & -      & \cmark &        & \cmark \\
    & DigiCert   & Username${}^1$                                          & \xmark & \xmark  & \cmark & \xmark & \cmark & \cmark & \xmark \\
    & Sectigo    & Email                                                  & \xmark & \xmark  & \cmark & \xmark & \cmark & \xmark & \cmark \\
    & GoDaddy    & Username${}^1$, customer No.${}^1$                      & \xmark & \xmark  & \cmark & -      & \cmark & \cmark & \cmark \\ %
    & GlobalSign & Username                                                & \xmark & \xmark  & \cmark & -      & \cmark & -      & \cmark \\
    \hline

\end{tabular}
\end{center}
\vspace{-10pt}
\centering \footnotesize -: No response. ${}^1$: Can be retrieved using domain name/Email.
\vspace{-8pt}
\caption{Password recovery at each provider.}
\label{password_recovery_table}
\end{table}

\subsection{Attacking Providers}
The adversary needs to poison the DNS cache of the provider, by injecting a record into the resolver's cache that maps the domain of the provider to the adversarial IP addresses. We therefore collect the IP addresses of the DNS resolvers of the providers. 

\subsubsection{Identify the target DNS resolvers}\label{sc:vic:rir}
In order to poison the DNS cache of the provider the adversary needs to find the IP addresses of the DNS resolvers which are used for looking up the Email servers of the customers during requests for password recovery.

We register accounts with the providers via the web portal of each provider. For our evaluation we register 20 accounts with each provider, each account is associated with a unique domain that we registered for that purpose. We use these registered accounts to learn about the infrastructure of the provider. We trigger the password recovery procedure for our registered accounts. To stay under the radar we limit the amount of password recovery requests to ten for each account. The Email server of the provider requests the DNS resolver to look up the \path{MX} and \path{A} records for our Email exchanger - this is required in order to send the password, or the link to reset the password. %
We monitor the requests arriving at the nameservers of our domains and collect the IP addresses which sent DNS requests for records in domains under which we registered our accounts. These IP addresses belong to the DNS resolvers used by the providers. We repeat this for each provider on our list in Table \ref{resolvers_accepting_fragments_table}. %

For every provider, we list in Table~\ref{resolvers_accepting_fragments_table} the service providers of the Email servers and the DNS resolvers (by mapping the observed IP addresses to ASNs). Additionally, we also performed measurements if the resolvers of the providers in our dataset support DNSSEC and the default EDNS size in DNS requests.

\subsubsection{Poison providers' DNS caches}\label{ssc:eval:poson}

To understand the vulnerabilities to cache poisoning across the providers, we evaluate the DNS cache poisoning methodologies against the DNS resolvers of the providers in our dataset. Our evaluations are done as described in Section \ref{sc:dns:intercept} using the victim domains that we set up and the accounts that we registered. During the evaluations the adversary triggers password recovery procedure and applies the DNS cache poisoning methodologies (one during each test) to inject into the DNS cache of the provider malicious records mapping the Email servers of our victim domains to the hosts controlled by the adversary.
In this section we report on the results of our evaluations and the extent of the vulnerabilities among the providers.

{\bf HijackDNS.} To infer the scope of providers vulnerable to the attack in Section \ref{sc:bgp:eval} we perform Internet measurements checking for vulnerabilities that allow sub-prefix hijacks. Since many networks filter BGP advertisements with prefixes more specific than /24, we consider an IP address vulnerable if it lies inside a network block whose advertised size is larger than /24. We therefore map all resolvers' IP addresses to network blocks. Then, to obtain insights about the sizes of the announced BGP prefixes for providers' network blocks with resolvers we use the BGPStream of CAIDA \cite{bgpstream} and retrieve the BGP updates and the routing data from the global BGP routing table from RIPE RIS \cite{riperis} and the RouteViews collectors \cite{routeviews}. We analyse the BGP announcements seen in public collectors for identifying networks vulnerable to sub-prefix hijacks by studying the advertised prefix sizes. The dataset used for the analysis of the vulnerable sub-prefixes was collected by us in January 2021. %
Our analysis in Table \ref{resolvers_accepting_fragments_table} shows that the networks of 29 providers are vulnerable to sub-prefix hijacks. %

 To understand the viability of same-prefix hijack attacks we perform experimental simulations using the target providers and customers in our dataset. For creating the topological map of the AS-relationship dataset of the customer domains and the providers in our dataset we use CAIDA \cite{caida}. We simulate the attacks using a simulator developed in \cite{hlavacek2020disco}. We evaluate \hijack\ attack for each provider with respect to customer domains of the corresponding provider and an AS level adversary on an Internet topology. In our simulations we consider attacks from 1000 randomly selected ASes against the domains of the customers and providers. The adversary can succeed at the attack against 80\% of the Alexa customer domains with 60\% success probability. One of the reasons for the high success probability is the concentration of the nameservers in few ASes: 10\% of the ASes host 80\% of the nameservers in Alexa domains and 1\% of ASes host 80\% of the domains. The customers of the LIRs are slightly more resilient since they mostly use at least two nameservers on different prefixes. This means that to succeed the attacker would need to hijack both. Furthermore, the distribution of the nameservers across ASes is more uniform in contrast to Alexa domains.

{\bf SadDNS \& FragDNS.} To test vulnerabilities in the providers to \sad\ and \frag\ we perform the evaluations in Sections \ref{sc:sad:eval} and \ref{sc:frag:eval}. Out of 31 tested providers, 28 (90\%) are vulnerable to \frag\ attack and four of the providers are vulnerable to \sad\ attack. Vulnerabilities for each provider are listed in Table \ref{password_recovery_table}.

\begin{table*}[!ht]
    \centering
    \footnotesize
    \begin{tabular}{|l|rl|c|rl|ccHH|ccccc|c|cc|cH|}
    \hline
    
 &  & Provider & \mr{2}{\makecell{Mail\\service\\provided\\by}} & \multicolumn{2}{c|}{Resolver} & \multicolumn{4}{c|}{Seen Via}  & \multicolumn{5}{c|}{Accept Fragment} & \multicolumn{1}{c|}{BGP} & \multicolumn{2}{c|}{DNSSEC} & \mr{2}{\makecell{EDNS\\size}} & Override \\ 
 & \rot{Rank} &  &  & \#  & \makecell[l]{service\\provided by} & \rot{Signup} & \rot{PW Rec.} & \rot{DV} & \rot{MX} & \rot{1500} & \rot{1280} & \rot{576} & \rot{292} & \rot{68} & \rot{\makecell{prefix-\\size}} & \rot{do} & \rot{validate} &  & \\ \hline
 \hline
 \multirow{5}{*}{\rot{RIRs}}
 & -   & AFRINIC          & Self        & 3   & Self         & \cmark & \cmark & -      & \xmark & \cmark & \cmark & \cmark & \cmark & \cmark &/23     & \cmark & \cmark & 4096    & NS-to-A \\ \cline{2-19}
 & -   & APNIC            & Self        & 1   & Self         & \cmark & \cmark & -      & \cmark & \cmark & \cmark & \cmark & \cmark & \cmark &/24     & \cmark & \cmark & 4096    & DNAME   \\ \cline{2-19}
 & -   & ARIN             & Self        & 4   & Self         & \cmark & \cmark & -      & \xmark & \cmark & \cmark & \cmark & \cmark & \cmark &/24     & \cmark & \cmark & 4096    & DNAME   \\ \cline{2-19}
 & -   & LACNIC           & Self        & 1   & Self         & \cmark & \cmark & -      & \cmark & \cmark & \cmark & \cmark & \cmark & \cmark &/22     & \cmark & \cmark & 1280    & DNAME   \\ \cline{2-19}
 & -   & RIPE NCC         & Self        & 3   & Self         & \cmark & \cmark & -      & \xmark & \cmark & \cmark & \cmark & \cmark & \cmark &/12-/23 & \cmark & \cmark & 4096    & DNAME   \\ \hline
 \hline
\multirow{11}{*}{\rot{Registrars }}
 & 1   & godaddy          & Self        & 3   & Self         & \cmark & \cmark & \xmark &        & \cmark & \cmark & \cmark & \cmark & \cmark &/19-/21 & \cmark & \xmark & 4096    & DNAME   \\ \cline{2-19}
 & 2   & namecheap        & SendGrid    & 64  & SendGrid     & \cmark & \cmark & -      &        & \xmark & \xmark & \cmark & \cmark & \cmark &/12-/23 & \cmark & \xmark & 1232    & \xmark  \\ \cline{2-19}
 & 3   & networksolutions & Self        & 1   & Self         & (3)    & \cmark & -      &        & -      & -      & -      & \cmark & \cmark &/20     & \xmark & (1)    & 512 (2) &         \\ \cline{2-19}
 & 6   & enom             & Self        & 17  & Self, Google & \cmark & \cmark & -      &        & \cmark & \cmark & \cmark & \cmark & \cmark &/20     & \cmark & \xmark & 4096    &         \\ \cline{2-19}
 & 9   & name.com         & Self (AWS)  & 8   & Self (AWS)   & \cmark & \cmark & -      &        & \cmark & \cmark & \cmark & \cmark & \cmark &/12     & \cmark & \xmark & 4096    &         \\ \cline{2-19}
 & 10  & Alibaba cloud    & Self        & 11  & Self         & \cmark & \cmark & -      &        & \cmark & \cmark & \cmark & \cmark & \cmark &/16-/21 & \cmark & \xmark & 4096    &         \\ \cline{2-19}
 & 11  & AWS              & Self        & 46  & Self         & \cmark & \cmark & -      &        & \cmark & \cmark & \cmark & \cmark & \cmark &/12-/21 & \cmark & \xmark & 4096    &         \\ \cline{2-19}
 & 12  & gandi            & Self        & 3   & Self         & \cmark & \cmark & -      &        & \cmark & \cmark & \cmark & \cmark & \cmark &/23     & \cmark & \cmark & 4096    &         \\ \cline{2-19}
 & 13  & namesilo         & Self        & 2   & Self         & \cmark & (1)    & -      &        & -      & -      & -      & \cmark & \cmark &/16-/19 & \xmark & \xmark & 512 (2) &         \\ \cline{2-19}
 & 14  & Google Cloud     & Self        & 120 & Self         & \cmark & (1)    & -      &        & -      & \cmark & \cmark & \xmark & \xmark &/16-/22 & \xmark & \xmark & 1232    &         \\ \cline{2-19}
 & 15  & OVHCloud         & Self        & 4   & Self         & \cmark & \cmark & -      &        & \cmark & \cmark & \cmark & \cmark & \cmark &/18-/24 & \cmark & \cmark & 4096    &         \\ \hline
 \hline
\multirow{14}{*}{\rot{IaaS Providers }}
 & 1   & Amazon AWS       & Self        & 46  & Self         & \cmark & \cmark & -      &        & \cmark & \cmark & \cmark & \cmark & \cmark &/12-21  & \cmark & \xmark & 4096    &         \\ \cline{2-19}
 & 2   & Microsoft Azure  & outlook.com & 373 & outlook.com  & \cmark & \cmark & -      &        & -      & -      & -      & \xmark & \xmark &/13-19  & \xmark & \xmark & 512 (2) &         \\ \cline{2-19}
 & 3   & Alibaba Cloud    & Self        & 11  & Self         & \cmark & \cmark & -      &        & \cmark & \cmark & \cmark & \cmark & \cmark &/16-/21 & \cmark & \xmark & 4096    &         \\ \cline{2-19}
 & 4   & Google Cloud     & Self        & 120 & Self         & \cmark & (1)    & -      &        & -      & \cmark & \cmark & \xmark & \xmark &/16-/22 & \xmark & \xmark & 1232    &         \\ \cline{2-19}
 & 5   & IBM Cloud        & SendGrid    & 51  & SendGrid     & \cmark & (1)    & -      &        & \xmark & \xmark & \cmark & \cmark & \cmark &/12-/23 & \cmark & \xmark & 1232    & \xmark  \\ \cline{2-19}
 & (7) & Tencent Cloud    & Self        & 13  & Self         & \cmark & \cmark & -      &        & \cmark & \cmark & \cmark & \cmark & \cmark &/12-/19 & \cmark & \xmark & 4096    &         \\ \cline{2-19}
 & (8) & Oracle Cloud     & Self        & 9   & Self         & \cmark & (1)    & -      &        & \xmark & \xmark & \xmark & \xmark & \xmark &/17-/23 & \cmark & \cmark & 1372    & DNAME, NS-to-A  \\ \cline{2-19}
 & -   & DigitalOcean     & Mailchimp   & 8   & Mailchimp    & \cmark & \cmark & -      &        & \cmark & \cmark & \cmark & \cmark & \cmark &/17-/22 & \cmark & \xmark & 4096    & \xmark  \\ \cline{2-19}
 & -   & Linode           & Self        & 2   & Self         & \cmark & \cmark & -      &        & \cmark & \cmark & \cmark & \cmark & \cmark &/17     & \cmark & \cmark & 4096    &         \\ \cline{2-19}
 & -   & IONOS            & Self        & 2   & Self         & \cmark & (1)    & -      &        & -      & -      & -      & \cmark & \cmark &/16     & \cmark & \cmark & 1220    &         \\ \cline{2-19}
 & -   & Hostwinds        & Postmark    & 15  & OpenDNS      & (3)    & \cmark & -      &        & \xmark & \xmark & \xmark & \xmark & \xmark &/19-/21 & \cmark & \cmark & 1410    & NS-to-A \\ \cline{2-19}
 & -   & OVHCloud         & Self        & 4   & Self         & \cmark & \cmark & -      &        & \cmark & \cmark & \cmark & \cmark & \cmark &/18-/24 & \cmark & \cmark & 4096    &         \\ \cline{2-19}
 & -   & Vultr            & Self        & 8   & Self         & \cmark & \cmark & -      &        & \cmark & \cmark & \cmark & \cmark & \cmark &/18-/20 & \cmark & \cmark & 4096    &         \\ \cline{2-19}
 & -   & CloudSigma       & Mailchimp   & 6   & Mailchimp    & \cmark & \cmark & -      &        & \cmark & \cmark & \cmark & \cmark & \cmark &/17-/22 & \cmark & \xmark & 4096    & \xmark  \\ \hline
 \hline
\multirow{5}{*}{\rot{CAs }}
 & 1   & IdenTrust        & Trend Micro & 114 & Trend Micro  & \cmark & (1)    & (3)    &        & \cmark & \cmark & \cmark & \cmark & \cmark &/15     & \cmark & (1)    & 4096    &         \\ \cline{2-19}
 & 2   & digicert.com     & Self        & 137 & Self         & \cmark & \cmark & (3)    &        & \cmark & \cmark & \cmark & \cmark & \cmark &/16-/22 & \cmark & (1)    & 4096    & DNAME   \\ \cline{2-19}
 & 3   & sectigo.com      & SendGrid    & 10  & SendGrid     & (3)    & \cmark & \xmark &        & \xmark & \xmark & \cmark & \cmark & \cmark &/12-/23 & \cmark & \xmark & 1232    & \xmark  \\ \cline{2-19}
 & 4   & godaddy          & Self        & 3   & Self         & \cmark & \cmark & \xmark &        & \cmark & \cmark & \cmark & \cmark & \cmark &/19-/21 & \cmark & \xmark & 4096    & DNAME   \\ \cline{2-19}
 & 5   & globalsign.com   & (1)         & 35  & Google       & \cmark & (1)    & \xmark &        & \cmark & \cmark & \cmark & \xmark & \xmark &/20     & \cmark & \cmark & 4096    &         \\ \hline
 
    \end{tabular}
    \caption{Measurement study of provider's DNS resolvers and Email servers. (1): Could not test. (2): No EDNS. (3): No Email after sign-up. -: Does not apply. }
    \label{resolvers_accepting_fragments_table}
\end{table*}

\subsection{Measurements of Vulnerable Customers}\label{sec:hitrate-study}\label{sec:hitrate-all}\label{sc:vuln-customers}

The success of the attack against a specific victim customer depends not only on the vulnerabilities in the DNS resolver of the provider but also on the properties in the domain of the customer. For instance, say a DNS resolver of some provider is vulnerable to \frag\ attack but the nameservers of the customer's domain do not fragment UDP packets and packets that are too large are transmitted over TCP. In that case, the \frag\ attack is not effective against that customer. To understand the extent of the vulnerabilities in customers we evaluate the attack methodologies in Section \ref{sc:dns:intercept} against the DNS resolvers that we own and control using the responses from the domains of the customers in our dataset. Using results from this evaluation we can reliably determine if the attack methodology is effective against a customer or not. %

The results of our experimental evaluations of attacks in Section \ref{sc:dns:intercept} and measurements of the customers' domains and their nameservers are summarised in Table \ref{rir_whois_database_details}.
\begin{table}[th!]
\begin{center}
\footnotesize
\setlength\tabcolsep{1.5pt}
\begin{tabular}{ |c|r|r|rH|Hrr|r|rr|}
    \hline
             & \mcc{2}{\# Resources}             & \mcc{2}{  }                     &          & \multicolumn{5}{c|}{Vulnerable to} \\
    \cline{2-3} \cline{6-11}
    Account  & \mr{2}{Total}  & found             & \# Acc- & domains           & \# Name- & \mcc{2}{BGP} & Sad- & \mcc{2}{FragDNS}\\ 
    Provider &                & e-mail            & ounts           &                   & servers  & sub & same     & DNS  & any  & global \\
    \hline \hline
    \multicolumn{5}{|c|}{\textbf{Scanned resources}} & \multicolumn{6}{c|}{\textbf{Vulnerable Accounts}}  \\
    \hline
\mr{2}{RIRs}   & \mr{2}{92,857} &  69,287  & \mr{2}{87,547} & \mr{2}{57,507} & \mr{2}{67,132}& 47,840  & n/a      & 8,469   & 14,136  & 1,193   \\   %
               &               & 75\% &               &               &               & 56\% & n/a      & 11\% & 17\% & 1.5\%  \\ 
\hline
Regis-         &\mr{2}{100,000} &  10,597  & \mr{2}{7,450}  &  \mr{2}{4,626} &  \mr{2}{8,177}& 3,308   & n/a      & 666     & 1,560   & 85      \\   %
trars          &               & 11\% &               &               &               & 45\% & n/a      & 10\%  & 21\% & 1.2\%  \\
\hline
\mr{2}{Both}   &\mr{2}{192,857} &  79,884  & \mr{2}{94,997} &  \mr{2}{ ?  } &  \mr{2}{ ?  } & 51,148  & n/a      & 9,135   & 15,696  & 1,278   \\   %
               &               & 41\% &               &               &               & 56\% & 80\%      & 11\% & 17\% & 1.4\%  \\
    \hline
    \hline
    
    \multicolumn{11}{|c|}{\textbf{Vulnerable resources}}  \\
    \hline

    \multicolumn{4}{|r}{IP Addresses}  & & & 81\%    & n/a & 30\%    & 51\% & 21\% \\ %
    \multicolumn{4}{|r}{AS Numbers}    & & & 60\%    & n/a & 12\%    & 20\% &  3\% \\ %
    \multicolumn{4}{|r}{Domains}       & & & 47\%    & n/a  & 10\%    & 27\% & 1\% \\ %
    
    \hline
    \hline
    \multicolumn{11}{|c|}{\textbf{Attack success probability}}  \\
    \hline
    
    \multicolumn{4}{|r}{Success probability} & & & 100\% & 60\% & 0.2\% & 0.1\% & 20\% \\
    
    \hline
    
\end{tabular}
\end{center}
\vspace{-10pt}
\caption{Customer-side vulnerability data}

\label{rir_whois_database_details}
\vspace{-10pt}
\end{table}

\begin{figure}
    \centering
    \includegraphics[width=0.48\textwidth]{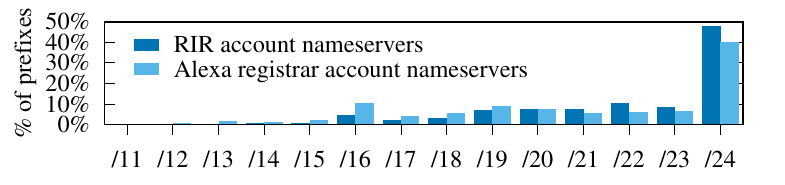}
    \vspace{-15pt}
    \caption{IP prefix distribution of customer accounts' nameservers}
    \label{fig:prefix_sizes}
\end{figure}

{\bf HijackDNS.} We analyse the prefixes of the customers similarly to Section \ref{ssc:eval:poson}. The results are plotted in Figure \ref{fig:prefix_sizes}. Our findings are that more than 60\% of the domains have all their nameservers on prefixes less than /24. Furthermore, above 20\% of the domains host all the nameservers of each domain on a single prefix, as a result, by hijacking one prefix the adversary immediately hijacks the entire domain. Out of these, 17\% host all the nameservers on a prefix that is less than /24. 10\% of the domains have a single nameserver in the domain. To conclude: more than 40\% of the domains are vulnerable to \hijack\ attack via sub-prefix hijack.

{\bf SadDNS.} Based on the implementation in Section \ref{sc:sad:eval} we develop an automated simulation of the \sad\ attack, and run it on our dataset of customer domains to compute the success probability of \sad\ attack against our victim DNS resolver. When running the attack for domains that have the required properties (e.g., support rate limiting), poisoning succeeds after an average of 471s (min 39s, max 779s) which is comparable to the original SadDNS results (avg 504s, min 13s, max 1404s, \cite{saddns}). Our test implementation triggers 497 queries on average for each domain, which is strongly correlated with the attack duration due to the fact that we do not trigger more than two queries per second; the resolvers return SRVFAIL when receiving more than two queries per attack iteration. By inverting this number we get a hitrate of 0.2\%.

Our results of an automated evaluation of \sad\ show that 8,469 accounts from the RIRs dataset and 11\% of the accounts from the Alexa dataset could be hijacked via the \sad\ method.

{\bf FragDNS.} We measured the victim customers' nameservers for support of ICMP errors and fragmentation. %
We send a DNS request to the nameserver for {\tt ANY} type DNS record. After a DNS response, we follow with an ICMP PTB error, that indicates different MTU values, and repeat the request. We check if the response arrived in fragments according to the MTU value indicated in the ICMP error message. We performed evaluations with the following MTU values of 1280, 576, 296 and 68 bytes.

 \begin{figure}
    \centering
    \includegraphics[width=0.48\textwidth]{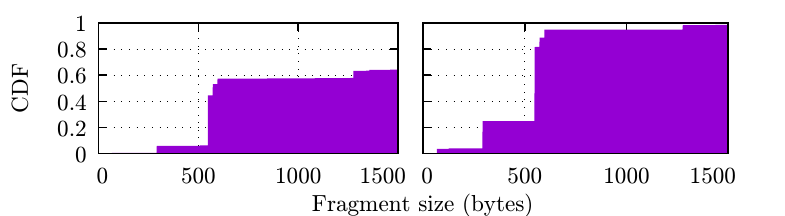}
    \caption{Cumulative distribution of lowest fragment size of nameservers (left) and domains (right) after sending ICMP PTB.}
    \label{fig:mtu_cdf}
    \vspace{-5pt}
\end{figure}
Figure \ref{fig:mtu_cdf} (Left) shows cumulative distribution of fragmented packet size we received after sending ICMP PTB. Right side shows the percentage of the domains where at least one nameserver supported a MTU smaller than the plotted size. As can be seen, for more than $90\%$ of the domains with PMTUD-configured nameservers, at least one nameserver is willing to reduce the fragment size (in response to ICMP PTB) to almost $548$ bytes, for roughly $35\%$ domains to $296$ bytes and for $10\%$ to 68 bytes. This essentially allows to inflict fragmentation to any size needed. Our evaluations in Section \ref{sc:frag:eval} indicate that 11964 RIR customer accounts (13.6\%) and 2352 Alexa domain holder accounts (22.2\%) are vulnerable to the \frag\ attack.

\begin{figure}[!t]
  \begin{center}
   \includegraphics[width=0.45\textwidth]{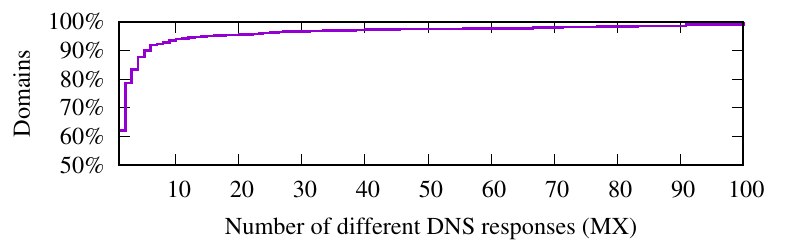}
   
\vspace{-5pt}
		\caption{CDF of number of observed DNS MX responses per customer email address domain (each nameserver was queried 100 times).}
		\label{fig:dns_reponses}
		\vspace{-10pt}
  \end{center}
\end{figure}

 \begin{algorithm}[!t]
\caption{Predictability of records in responses.}
\label{algo:udp-checksum}
\SetAlgoLined
\scriptsize
\For {each (domain, nameserver)} {
    initialise set of different DNS responses as empty\;
    \For {$batch=1,2,\ldots,25$} {
        \For {$iteration=1,2,3,4$} {
            send the same DNS request\;
            \If {new response arrived} {
                add the new response to the response set\;
            }
        }
        \If {no new responses in last batch} {
            break\;
        }
    }
    record number of different DNS responses\;
}
\end{algorithm} 
We analyse the attacker's success probability of crafting the spoofed second fragment with the correct UDP checksum and the correct IPID value. To compute the success rate to hit the correct UDP checksum we performed the following evaluation. For each customer domain in our dataset, we query the nameservers of each domain multiple times sending the same DNS request (with the domain of the customer's email and type MX), and check if the DNS responses from the nameservers contain the same DNS records and the same order of DNS records, during each iteration. The computation of the UDP checksum for each domain is described in pseudocode in Algorithm \ref{algo:udp-checksum}. Our evaluation shows that for 1748 domains ($62\%$), nameservers always return the same DNS response (with the same records and sorted in the same order); see Figure \ref{fig:dns_reponses}.
For our measurement of the IPID allocation methods supported by the nameservers of the customers we use the following methodology. We issue queries from two hosts (with different IP addresses). Data per nameserver is listed in Table \ref{tab:ipid-allocation}. Our measurements show that 290 vulnerable nameservers ($4.88\%$) use a globally incremental IPID assignment. The computation of the IPID allocation for each domain are described in pseudocode in Algorithm \ref{algo:ipid-allocation}.

\begin{algorithm}[ht!]
\caption{IPID allocation in nameservers.}
\label{algo:ipid-allocation}
\SetAlgoLined
\scriptsize
\For {each (domain, nameserver)} {
    \For {$batch=1,2,3,4$} {
        send DNS request from $Prober_1$\;
        record IPID in DNS response as $IPID_{2*i-1}$\;
        send DNS request from $Prober_2$\;
        record IPID in DNS response as $IPID_{2*i}$\;
    }
    \If {$IPID_i, i=1,2,\ldots,8$ is incrementing} {
        globally incrementing\;
    }
    \If {$IPID_i, i=1,3,5,7$ or $IPID_i, i=2,4,6,8$ is incrementing} {
        per-dest incrementing\;
    }
    \eIf {$IPID_i==0, i=1,2,\ldots,8$} {
        zero\;
    } {
        random and other\;
    }
}
\end{algorithm}

\begin{table}[bht!]
    \vspace{-7pt}
    \centering
    \footnotesize
\setlength\tabcolsep{3.2pt}
    \begin{tabular}{|c|c|c|c|c|c|c|}
    \hline
    &          &         &         & Random           &          &       \\
    & Per-Dest & Global  & Zero    & and other        & N/A      & Total \\
    \hline
    \multirow{2}{*}{All}
    & 64.58\%  & 8.31\%  & 4.89\%  & 11.92\%          & 10.30\%  & 100\% \\
    & 45308    & 5829    & 3434    & 8364             & 7223     & 70158 \\
    \hline
    \multirow{2}{*}{Frag}
    & 53.96\%  & 4.88\%  & 13.75\% & 23.67\%          & 3.74\%  & 100\% \\
    & 3206     & 290     & 817     & 1406             & 222     & 5941  \\
    
    \hline
    \end{tabular}
    \caption{IPID allocation of all nameservers and of fragmenting nameservers.}
    \label{tab:ipid-allocation}
\end{table}

\begin{figure}[!t]
  \begin{center}
   \includegraphics[width=0.45\textwidth]{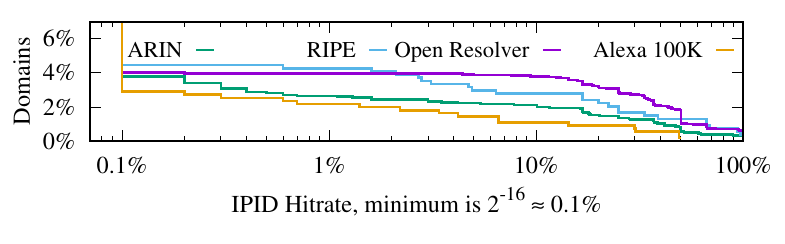}
   
        \vspace{-10pt}
		\caption{Reverse CDF to correctly guess the IPID for all customers' domains.}
		\label{fig:hitrate_ipid}
  \vspace{-20pt}
  \end{center}
\end{figure}

We automate the attack in Section \ref{sc:frag:eval} and execute the entire \frag\ attack against all the vulnerable customer domains, by injecting malicious records mapping Email servers of customers to an IP address of our adversarial host. Our evaluation combines the data we collected on DNS records in responses (randomisation of the DNS records or of their order in responses) and the IPID allocation of the nameservers. We also used Algorithm~\ref{algo:ipid-allocation} to estimate the IPID increment rate, by recording the timestamp of each response and calculating the average increment rate of IPID value. %
We then extrapolate the value of IPID and calculate the probability of our adversary to correctly place at least one out of 64 fragments\footnote{64 fragments is the minimal size of the IP-defragmentation cache.} with the matching IPID in the resolver's defragmentation cache. We use different values for IPID increment rate and delay between the query, which probes the IPID value, and the IPID value that was de-facto assigned to the DNS response by the nameserver. Results are plotted in Figure \ref{fig:hitrate_ipid}. For example, the IPID prediction success rate is over 10\% for roughly 3\% of RIPE, 2\% of ARIN and 1\% of 100K-top Alexa customers. Success rates for ARIN and 100K-top Alexa customers are lower mostly because of the higher latencies of those, see Figure \ref{fig:latency}. For nameservers which do not use globally incremental IPID, we assume a hitrate of $64 / 2^{16}$ which is achieved by just randomly guessing the IPID. %

The probability to compute the correct checksum is capped at a minimum of $1/2^{16}$ in case of nameservers which generate responses with different records or with random ordering of records. Finally the probabilities to correctly compute both, the IPID value and the order of records to get the correct UDP checksum, are multiplied resulting in the combined hitrate. 

Our automated attack against all the customers shows that around $2\%$ of the domains (5 for RIPE, 17 for ARIN) have a success probability higher than $10\%$. Furthermore, for about $20\%$ of the domains, success probability is over 0.1\% which is a consequence of non-predictable IPID allocation and the stable DNS records in responses generated by these domains. When the DNS response can be predicted, even with a random IPID allocation method, an attacker has a hitrate of about $64/2^{16} \approx 0.1\%$. At this hitrate, when the attacker performs the attack multiple times, the probability to conduct the attack successfully at least once is $50\%$ at around 700 repetitions.

Our automated evaluation provides a lower bound for successful attacks against a randomly chosen domain -- this is a worst case analysis since it also considers domains which are much more difficult to attack, e.g., since they use servers with random IPID allocation, servers with high traffic rate, and servers which return different number and order of records in responses. Adjusting the attack parameters manually against a given victim customer domain results in a much higher attack rate. Furthermore, against many customer domains with low traffic volume, incremental IPID values and fixed number and order of DNS records, the attacker can reach above 90\% success rate.

\begin{table}[!ht]
\begin{center}
\footnotesize
\setlength\tabcolsep{2.0pt}
\begin{tabular}{| c | c|c | cccc | c |}
    \hline
    \rot{ \makecell{ ~Additonal~ \\ ~Validation~ } } & \mcc{2}{Attack} & \rot{RIRs} & \rot{Registrars} & \rot{IaaS} & \rot{CAs} & \makecell{Outcome / \\ Attacker use } \\
    \hline
    \hline
    
    RIRs   & \mcc{2}{Account transfer/delegation}   & \cmark & \cmark & \cmark & \xmark & permanent control \\ \hline
    No     & \mcc{2}{Changing the account details}  & \cmark & \cmark & \cmark & \cmark & permanent control \\ \hline
    RIRs   & \mcc{2}{Close the account permanently} & \cmark & \cmark & \cmark & \cmark & DoS               \\ \hline
    No     & \mcc{2}{Disabling Email alerts}       & \cmark & \cmark$^*$ & \xmark & \cmark$^*$ & remain stealthy   \\ \hline
    \hline
    
    \mr{2}{~RIRs} & \mcc{2}{\mr{2}{Resource transfer}}&\cmark & \cmark & \cmark & \xmark & permanent control \\ \cline{4-8}
           & \mcc{2}{}                              & \cmark & \cmark & \xmark & \xmark & sell resources    \\ \hline
    No     & \mcc{2}{Resource return / deletion}    & \cmark & \cmark & \cmark & \cmark & DoS               \\ \hline
    \mr{2}{CAs}&\mcc{2}{\mr{2}{Purchase new resources}} &\cmark&\cmark& \cmark & \cmark & financial Damage  \\ \cline{4-8}
           & \mcc{2}{}                              & \cmark & \cmark & \cmark & \cmark & anonymous usage   \\ \hline
           
    \mr{3}{~No} & \mr{3}{\makecell{Control / Modify \\ Resources}} 
            & Whois DB                              & \cmark & \cmark & \xmark & \xmark & facilitates hijacking \\ \cline{4-8}
           && VMs                                   & \xmark & \xmark & \cmark & \xmark & various           \\ \cline{4-8}
           && NS records                            & \xmark & \cmark & \xmark & \xmark & traffic hijacking         \\ \hline
    \hline
    
    No     & \mcc{2}{Create new ROAs/certificates}  & \cmark & \xmark & \xmark & \cmark & facilitates hijacking \\ \hline
    No     & \mcc{2}{Create invalid ROAs}           & \cmark & \xmark & \xmark & \xmark & DoS \\ \hline
    No     & \mcc{2}{Revoke certificates}           & \xmark & \xmark & \xmark & \cmark & DoS \\ \hline

\end{tabular}
\end{center}
\vspace{-10pt}
\caption{Actions an attacker can carry out after hijacking a customer's account. $^*$The Email address where the alert is sent can be changed. Additional validation requires either that additional documents are sent, or in case of issuing a new TLS certificate, that a domain validation must be passed.}%
\label{possible_exp}
\vspace{-10pt}
\end{table}
\section{Manipulation of Digital Resources}\label{sc:exploits}

In this section we demonstrate exploits that the adversary can perform when controlling an account of a (victim) customer. Most of the actions are similar across the providers, even providers of different infrastructure, such as RIRs and the domain registrars. Hence, we in details explain our demonstration of the exploits by taking over a victim LIR account, and then briefly describe the exploits we evaluated by taking over our victim accounts with the other providers. For our demonstration we select RIPE NCC RIR, GoDaddy domain registrar, Microsoft Azure IaaS provider and DigiCert CA. In order to evaluate the exploits using an account of a network operator, we cooperate with a large customer under RIPE NCC. We cooperate with that LIR and use a real account that has an {\em operator}/{\em administrator} role\footnote{RIPE NCC Single Sign-On (SSO) accounts are general authentication mechanism for all web-based services provided by RIPE NCC that include customer portal and other harmless services - for example RIPE Meeting facilitation.}. For domain registrars, IaaS and CAs, we used our own accounts which were used to buy test resources to test the possibilities of the account. We summarise the exploits for different providers in Table~\ref{possible_exp}.

For our evaluation of the exploits, we first carry out our attacks in Section \ref{sc:dns:intercept} to take over the victim accounts, and then carry out the exploits. We do this in order to understand what notifications are sent to the genuine account owners during such attacks, what actions can be performed, and which are prevented.

\subsection{Regional Internet Registries}
\label{sec:exploits:rirs}
We show that adversary, controlling an SSO account of a victim LIR, can manipulate all the Internet resources associated with that LIR, e.g., the IPv4 and IPv6 addresses, ASNs, reverse DNS names to IP addresses mappings. The amount of resources managed by LIRs can vary enormously. There are small LIRs that manage just own AS, one IPv6 prefix and one or a very few IPv4 prefixes of minimal allocation size. %
There are also large LIRs managing vast PA address pools (Provider Aggregatable Addresses) for multiple clients. For instance, RIPE NCC has cumulative allocation of 587 202 560 IPv4 addresses. %

The adversary impersonating an administrator with the RIPE NCC SSO account holder can initiate different actions that lead to disruption or degradation of the services that are tied to the IP resources managed by the victim LIR. The adversary can even initiate transfer of IPv4 addresses that belong to the victim LIR to obtain direct financial benefit from that process. Our experimental evaluation with an SSO account under RIPE NCC RIR shows that the actions of the attacker do not trigger alerts and can be detected when the LIRs realises that its digital resources are gone. The access to RIPE NCC SSO account with {\em operator} or {\em administrator} roles for the victim's LIR opens to a range of possible exploits. %
We explain selected exploits with an example victim LIR under RIPE NCC below (also summarised in Table \ref{possible_exp}); the attacks similarly apply to the other RIRs.

{\bf RPKI administration.} Attacker creates/deletes/modifies Route Origin Authorizations (ROAs) in hosted RPKI system. This has two purposes: (1) to disrupt the propagation of the legitimate BGP updates of the resources managed by the victim LIR and (2) to facilitate BGP hijacking by authorising attacker's ASN to originate any subset of IP prefixes that are managed by the victim LIR. Networks which have deployed RPKI and perform filtering of BGP announcements with ROV will not trigger any alerts when the attacker issues a BGP announcement for a sub-prefix with a valid (yet fraudulent) ROA. %
    We consider creation of ROA with origin set to ASN0 ("always drop" as per \rfc{7607}) for a specific prefix within the resource pool managed by the victim LIR to be a special case of the malicious ROA intended to disrupt routing and cause DoS for the services tied to the IP addresses in question. Our measurements found that currently the Route-Origin Validation (ROV) is far from being universally deployed, with only 2190 ASes filtering invalid BGP announcements. Nevertheless, this is an increase in contrast to measurements from a few years ago, which found 71 ASes to validate ROV \cite{DBLP:conf/dsn/HlavacekHSW18,DBLP:journals/corr/0002BCKSW17}. Even with 2000 validating ASes, this type of attack is likely to cause only minor disruption in service availability and will remain unnoticed for extended period of time. %

{\bf RIPE DB modifications}. Attacker manipulates records in RIPE DB - the Internet addressing resource registry of the region and Internet Routing Registry (IRR) in one converged database. Modification of records in resource registry allows impersonation of the victim LIR's representatives in order to transfer resources from the victim LIR to unsuspecting recipient. %

IRR records are prerequisite for BGP hijacking attacks, because without proper records in IRR the attacker would not be able to persuade any well-managed upstream provider that is consistent with AS operation best practices to accept the fraudulent BGP announcements in the input filters on the BGP sessions. Attackers without the ability to modify IRR have to use less vigilant and generally poorly managed networks as upstream providers or have to utilise path manipulation attacks - both restrict success rate and stealthiness of the attack.

Creating the IRR records contradictory to the state in BGP is a way to partially disrupt route propagation and thus traffic. It can also de-stabilise network and significantly complicate network operation for the legitimate administrators. Route servers in majority of Internet Exchanges and major networks use IRR data in automatically generated filters that are applied on incoming BGP announcements. As a result of the contradicting IRR records these networks will drop or de-prefer the announcements from the legitimate resource holder. Moreover, well-managed networks keep manually generated import filters on small-scale BGP sessions for both peering and downstream customers. When a new session is set up or when a new prefix is about to be propagated from the neighbouring AS that is subject to filtering, the administrators manually check the IRR and resource registry to verify that the announcement is legitimate. Failing to have the proper records in IRR and in resource registry leads to refused BGP peerings, excessively strict BGP filters and therefore to dropped routes and overall degradation of the Internet connectivity for the victim network.

\begin{figure}[!ht]
    \centering
      \vspace{-10pt}
  \includegraphics[width=0.47\textwidth]{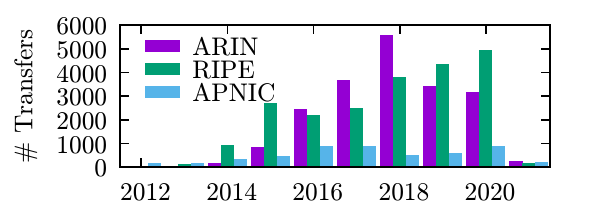}
  \vspace{-10pt}
   \caption{IPv4 resource transfers per year (PI \& PA).}
    \label{fig:ip:xfer}
    \vspace{-5pt}
\end{figure}

{\bf Initiating IPv4 addressing resource transfer}. Attacker sells the resources managed by the victim LIR. The potential gain from successfully completed attacks of this type is determined by the amount of the addresses managed by the LIR and the expected monetary value of IPv4 addressing resources. %
After the IPv4 regular pool depletion on 4 September 2012 each LIR is eligible for allocation of a single {\em /22} (1024 IPv4 addresses) prefix as per current IPv4 Address Allocation and Assignment Policies for the RIPE NCC Service Region (ripe-720) \cite{ripe-720}. We performed IPv4 /22 prefix transfer of an LIR under RIPE NCC which has not triggered alerts. This is not surprising since the IPv4 addresses' transfer is performed regularly by the RIRs, e.g., RIPE NCC and ARIN perform thousands of transfers per year, see statistics on IP transfer (PI and PA) we collected from the RIRs in Figure \ref{fig:ip:xfer}. %

To sell the IPv4 addresses belonging to the victim LIR, we needed to perform the following:

(1) modify the relevant LIR contacts (Emails, phone numbers) in the RIR database and the IRR to receive the communication from the RIR intended to the genuine LIR, the buyer as well as the other parties relevant to the IP resource transfer and prevent the victim to learn about the resource transfer process. This is performed via the victim SSO account which the attacker took over.

(2) find the buyer for the resources and to impersonate the victim representatives to successfully close an IPv4 resource transfer contract; we used the compromised SSO account to sell and transfer the resources to an LIR that we set up for that purpose. In practice, the attacker can also collude with a malicious adversary, which will perform the transaction and afterwards will legally own the resources. The victim will need to prove to not have authorised the payment. In fact the resolution outcome of such a case is not clear since the RIRs have not faced this attack before. 

(3) release the IPv4 prefix from the BGP routing table. This needs to be done so that the buyer believes that the resources are free and are being legitimately sold. This is done by sending an Email (via the Email address that the attacker modified in IRR and registry DB) to the upstream provider of the victim. The list of the upstream providers is available and can be obtained from the IRR. In the Email the attacker instructs the upstream provider to update the input filters and drop the network (that the attacker wishes to sell). Such requests are common, and do not require authentication or verification of requester's identity (e.g., with PGP/PKI) and the {\em sole source of truth that is checked prior applying the filters is RIPE DB} which, as we mentioned, the attacker can update via the SSO account.

Notice that if the buyer is colluding with the attacker - this step is not needed.

(4) Email a scanned IPv4 resource transfer contract to RIPE NCC. The contract has to match the company details of the victim and contain all formalities and certifications appropriate to the legal system in which the contract has been made. Moreover the contract has to be supplemented by extract from chamber of commerce or appropriate commercial registry that makes it possible to establish that the contract is signed by the eligible persons on both sides. This is simple to forge - the attacker can find and copy the signatures of the owner LIR online and paste them into the scanned contract document that the attacker prepares.

Extract from chamber of commerce is also simple to forge - for instance, in Czech Republic (CZE) the attacker has to go to any post office, pay 1 EUR and get either paper version with a stamp or PDF with PKI signature of state-operated CA. In any case, since the attacker only needs to send a scanned version of the document to RIPE, the attacker can get the document for any company and adjust it using Photoshop and it is accepted.

Finally, notice that the RIRs have limited personnel which, depending on the RIR, may need to deal with tens of transfers per day, see Figure \ref{fig:ip:xfer}. As a result, the adversary may often manage to sell the resources without raising alarms even when not satisfying these simple four steps. For instance, RIPE NCC has just 24 employees responsible for IP address distribution\footnote{\url{https://www.ripe.net/about-us/staff/structure/registration-services}} and there are more than 20 transfers per day, see Figure \ref{fig:ip:xfer}.

{\bf User and role management}. Attacker that controls an account with {\em administrator} role assigns other newly created users either {\em operator} or {\em administrator} roles for the victim LIR. This effectively hides the activities of the attacker for long periods of time even though the legitimate holder is actively operating the LIR. %

{\bf Modification of the LIR contacts and details}. There are two sets of Email and postal addresses and phone numbers related to the LIR - the first set is published in IRR and RIPE DB and it is tied to the resources and published to facilitate operation of the network and solving technical problems and those contacts are also used by the RIPE DB software itself for generating notifications about changes in RIPE DB. The second set contains the designated LIR contact information, namely the primary point of contact for the LIR and a contact for billing-related matters. Moreover, there is a postal address, that may differ from legal address of the LIR company and it can be modified in LIR portal. The attacker can redirect or change the LIR contact information to avoid detection by the victim LIR staff when activities that result in notifications or follow-up Emails are to be executed. Modifying LIR contacts will also make any attempt to rectify damages caused by the attack, when detected, harder.%

{\bf Termination of LIR membership.} Attacker initiates termination of LIR membership with the RIPE NCC by submitting a forged termination request via a written notice sent by Email. Forgery cases are not new and have already been seen in the past\footnote{\url{https://mailman.nanog.org/pipermail/nanog/2011-August/039379.html}}.

{\bf Modification of LIR organisation name, legal address and VAT number}. The attacker steals the LIR and all its IP addressing resources by pretending a transfer of ownership to other company. A scan of (the forged) contract of company acquisition has to be attached to the request in electronic form. %

{\bf Requesting new or voluntarily returning IP addressing resources}. The attacker requests or returns IP addressing resources from or to RIPE NCC. If the LIR is eligible for allocation of any scarce resources, the attacker obtains a new IP prefix that is not used in default-free zone (DFZ) and thus fulfils the prerequisite for transfer of not being announced. However, according to the current policy the newly obtained allocation of scarce resources (IPv4 addresses, 16-bit ASN) can not be transferred within 24 months from the allocation. The attacker can nonetheless hijack the resources for own purposes immediately and attempt to sell the resources after the grace period.

\subsection{Domain Registrars}
\label{sec:exploits:registrars}

Domain registrars handle the ownership of domains in name of the customer. We map the 100K-top Alexa domains to registrars with whom these domains are registered in Table \ref{password_recovery_table}. We demonstrate exploits that the attacker can perform when taking over an account with GoDaddy\footnote{The other domain registrars allow similar actions.}. The adversary can change the nameservers' IP addresses, which allows it to hijack the victim customer domain. This can be exploited to redirect clients to phishing websites. The adversary can delegate account access to itself or perform intra-and inter-registrar\footnote{It takes 60 days for an inter-registrar transfer to finalise.} domain transfer. The adversary can change the Email forwarding settings of the account which would allow it to hijack the Emails forwarded to the owner of the Email address. The adversary can also delete the domains associated with the compromised account and even close the account. The account owners can enable two-factor authentication for manipulation of resources associated with their account. However, this is not enabled by default, and is up to each customer to enable it. %

\subsection{Infrastructure as a Service}
\label{sec:exploits:iaas}

We evaluated the exploits that the adversary can carry out on resources associated with accounts at cloud providers. The adversary can manipulate virtual resources associated with the account, such as virtual machines, network interfaces, disk. The adversary can also exploit these resources to carry out attacks against victims in the Internet or victims located on the same cloud platform, e.g., via side channels \cite{ristenpart2009hey}. In addition the adversary can create new accounts with owner privileges or transfer subscription to another Azure account. 
\subsection{Certificate Authorities}
\label{sec:exploits:cas}

When controlling an account of a customer with a CA the adversary can revoke certificates and reissue existing certificates that were issued under that account. This allows to change the key-pair associated with a certificate. Nevertheless, some CAs, do not enforce any validation on reissuing certificates, in contrast to validation (domain validation, organisation validation or extended validation) that is enforced when requesting to issue a certificate for the first time. Therefore, instead of attacking domain validation procedure of the CAs it is more profitable to hijack a victim customer account and change the keys associated with the certificates for domains that the adversary targets. DigiCert and GoDaddy do not perform additional validation on requests to reissue certificates. Sectigo, GlobalSign, IdenTrust validate all requests to reissue certificates.

\section{Vulnerable Digital Resources}\label{sc:digital:resources}

The large fraction of the accounts under different providers that can be hijacked is alarming. Even more disturbing are the exploits that the adversaries can do with the resources assigned to the accounts. {\em What is the extent of the Internet resources that are at immediate risk due to the vulnerable providers and vulnerable customers?}

To answer this question we perform a correlation between the accounts that our study found to be vulnerable to hijacks via any of the attack methodologies in Section \ref{sc:dns:intercept} and the digital resources (domain names, IP addresses and ASNs) that the attacker can take over as a result of hijacking that account. In our analysis we consider only the domain registrars and the RIRs and their customers. Since there is no public database of customers of cloud providers and certificate authorities we exclude them from this analysis\footnote{These can be collected via a dictionary attack against the provider: the adversary inputs usernames and checks for error messages. Such a study however creates a significant load on the infrastructure of the provider.}. We list the correlation between the vulnerable resources and the vulnerable accounts in Table~\ref{tab:vulnerable_resources}.

\textbf{IP resources.} We compute the fraction of the assigned AS Numbers (ASNs) as well as assigned IPv4 address space which could be taken over by hijacking the vulnerable accounts of customers of RIRs. For this purpose, we combine IP-to-ASN and ASN-to-LIR mappings with our customer vulnerability data, which allows us to evaluate vulnerabilities in 73\% of the assigned IPv4 address space (for 27\% we could not extract LIR account information). Our results show that using any of the attack methodology in Section \ref{sc:dns:intercept} the adversary could take over 68\% of the IPv4 address space. This constitutes 93\% of the address space assigned to the accounts in our dataset. Even the weaker attack methodologies (\frag\ and \sad), which do not require controlling a BGP router, allow the adversaries to take over 59\% of the address space. Similarly, 74\% of the ASNs are associated with the accounts that can be hijacked via any of the DNS cache poisoning attacks in Section \ref{sc:dns:intercept} and 30\% with the \sad{} or \frag{} attack. The difference between the vulnerability volume for IP addresses and ASNs is due to the fact that large parts of the IPv4 address space is owned by a small number of ASes, e.g., %
21\% of the assigned IPv4 address space is attributed to the top 10 LIR accounts. %

\textbf{Domain resources.} We use our domain-to-account mapping to determine user accounts at registrars. This includes 11\% of the accounts for which we were able to extract customer account information. We believe however that the fraction of the vulnerable accounts is representative of all the 1M-top Alexa domains since the vulnerabilities only depend on the nameservers of the customers' domains. Our study shows that 65\% of the domains could be hijacked via any of \hijack{}, \sad{} or \frag{}, while 35\% could be hijacked via \sad{} or \frag{}.
\begin{table}
    \centering
    \footnotesize
    \setlength\tabcolsep{3.5pt} 
    \begin{tabular}{|r|c|c|c|c|c|}
    \hline
                  & \hijack{}  & \sad{}     & \frag{}   & Any     & \makecell{\sad{} or\\ \frag{}} \\
                  \hline
    IP addresses  & 81\%    & 30\%    & 51\%   & 93\% & 59\% \\
    Domains       & 47\%    & 10\%    & 27\%   & 65\% & 35\% \\
    \hline
    \end{tabular}
    \caption{Vulnerable resources mapped to accounts in our dataset.}
    \label{tab:vulnerable_resources}
\end{table}

\section{Recommendations for Countermeasures}
\label{sc:mitigations}

The fundamental problem that our attacks outline is the stealthiness and ease with which the adversaries can apply changes and manipulations over the Internet resources of providers of digital resources. %
Since Internet resources form the foundations for the stability and security of democratic societies, our work calls for a revision of the current practices of resource management and development of techniques that would secure the transactions over the Internet resources. For instance, selling Internet blocks should not happen immediately, and should require more than merely a scanned document over Email (which is easy to fake). In addition to the standard recommendations for hardening the DNS caches or blocking ICMP error messages, which we summarised in Table \ref{tab:countermeasures}, we also provide recommendations for best practices for providers and customers. %
\begin{table}[!t]
    \footnotesize
    \centering
    \setlength\tabcolsep{2.6pt}
    \begin{tabular}{|Hl|c|c|cccHHHH|}
    \hline

    & Countermeasure & Layer & \makecell{Provider- / \\ Customer- \\ side} & \rot{FragDNS} & \rot{SaDDNS} & \rot{HijackDNS} & \rot{on-path Email (BGP)} & \rot{password phishing} & \rot{on-device Malware} & Effect \\
    
    \hline
    
    & 2-FA TAN with out-of-band notif. & Web portal & both$^1$ & \cmark & \cmark & \cmark & \cmark & \cmark & \cmark & authenticates each transaction out-of band \\
    & 2-FA login                             & Web portal & provider & \cmark & \cmark & \cmark & \cmark & \cmark & \xmark & authenticates login out-of-band \\
    & IP-level account access restrictions   & Web portal & both     & \cmark & \cmark & \cmark & \cmark & \cmark & \xmark & prevents remote attacker to log-in \\
    & DNSSEC signing and validation          & DNS        & both     & \cmark & \cmark & \cmark & \xmark & \xmark & \xmark & prevents DNS record modifications \\
    & Disable/Patch ICMP rate-limit          & IP         & provider & \xmark & \cmark & \xmark & \xmark & \xmark & \xmark & prevents SadDNS attack \\
    & Disable NS rate-limit                  & DNS        & customer & \xmark & \cmark & \xmark & \xmark & \xmark & \xmark & prevents SadDNS attack \\
    & Disable PMTUD                          & IP         & customer & \cmark & \xmark & \xmark & \xmark & \xmark & \xmark & prevents fragmentation attack \\
    & Blocking Fragments                     & IP         & provider & \cmark & \xmark & \xmark & \xmark & \xmark & \xmark & prevents fragmentation attack \\
    
    \hline

    & MTA-STS \rfc{8461}                      & Email      & both     & \cmark & \cmark & \cmark & \cmark & \xmark & \xmark & encrypts password recovery Email \\
    
    \hline

    & Hide public account details            & General    & both     & \cmark & \cmark & \cmark & \cmark & \cmark & \xmark & complicates finding of account details \\
    & Request rate-limiting                  & Web portal & provider & \cmark & \cmark & \xmark & \xmark & \xmark & \xmark & complicates triggering DNS queries \\
    & Captchas                               & Web portal & provider & \cmark & \cmark & \xmark & \xmark & \xmark & \xmark & complicates triggering DNS queries \\
    & Separate systems                       & Web portal & provider & \cmark & \cmark & \xmark & \xmark & \xmark & \xmark & complicates triggering DNS queries \\
    & Resolver hardening                     & DNS        & provider & \cmark & \cmark & \xmark & \xmark & \xmark & \xmark & complicates off-path poisoning \\
    & Non-predictable IPID increment         & IP         & customer & \cmark & \xmark & \xmark & \xmark & \xmark & \xmark & complicates off-path fragment injection \\
    
    \hline
    
    & Out-of-band notifications              & Web portal & provider & \cmark & \cmark & \cmark & \cmark & \cmark & \cmark & notifies the victim upon account usage \\
    
    \hline
    \end{tabular}
    \caption{Countermeasures against different types of attackers. $^1$: requires the user to verify the out-of-band delivered transaction details before entering the TAN.}%
    \vspace{-10pt}
    \label{tab:countermeasures}
\end{table}

{\bf Separate system for high privileged users.} Currently, any user can create an account with most of the providers. The accounts can be used for managing Internet resources (high privileged) as well as for registering for events or mailing lists (low privileged). Low privileged accounts in the user management system have access to the same infrastructure (Email servers, DNS resolvers, etc) as the high privileged accounts, such as those of network operators. This enables adversaries to open low privileged accounts and use them to collect information about the infrastructure of the provider. The providers should use separate user management systems and a separate set of servers for users which own digital resources vs. users that, e.g., are registered to mailing lists or events.

{\bf Two-factor authentication.} Two factor authentication (2-FA) systems must be enabled by default. The two authentication factors must be independent of each other and an attacker should not be able to compromise both factors within a single attack. This for instance, rules out Email-based 2-FA for password recovery which is available at some of the providers we tested. %

{\bf Deploy captchas.} Our study shows that most providers do not use captchas, e.g., three out of five RIRs do not use captchas. Although captchas do not prevent the attack, they force the attacker to run manual tests making the attack more expensive to launch. Resolving the captchas is tedious and burdensome for the attacker (as well as for the researchers) to carry out in contrast to automated study of the victims. For instance, for studying vulnerabilities in DNS caches and for performing cache poisoning attacks we needed to run multiple password recovery procedures for triggering DNS requests to our domain. This study could not be automated in RIRs that use captchas.

{\bf Notifications of modifications.} Changes performed over the resources of providers either do not generate any notifications or generate notifications to the Email configured in the compromised account. First, the Email notifications will be received by our adversary, since it hijacked the victim domains in the resolver of the provider, and second, the adversary can change the contact Email in the account, and even disable notifications. The accounts with providers should be associated with contact Email which cannot be changed through the account and which is different than the one used to access the account.

{\bf Email address masking.} The Email addresses in the {\tt whois} records of the domains should be masked. Some of the domain registrars are already following this practice.

{\bf Account level IP address access restriction.} The registrars should restrict account access to only few static IP addresses belonging to the domain's owner.

{\bf Deploy DNSSEC.} DNSSEC (\rfc{4033} to \rfc{4035}) would essentially make the attack methodologies in Section \ref{sc:dns:intercept} practically impossible. %
Unfortunately, only 3.78\% domains of customers of RIRs and 5.88\% domains of customers of registrars are correctly signed. For instance, out of 1832 LIR domains under AFRINIC, only 58 are signed, and 27 of these domains are still vulnerable since the DNS resolvers cannot establish a chain of trust to them from the root anchor. Further, 12 use weak cryptographic keys (below 512 bits) and 12 use weak (vulnerable) hash functions. The remaining 95 domains out of those 1832 domains were not responsive. %
Unfortunately, even when the domain is signed and the resolver validates DNSSEC, as long as the human factor is in the loop, there is risk for vulnerabilities and misconfigurations, \cite{shulman2017one,chung2017longitudinal}. Hence we recommend that the providers and customers deploy additional measures that we list in this section to harden their infrastructure.

\section{Conclusions}\label{sc:conclusion}

Each provider maintains a database that defines which customer owns which Internet resources and offers tools for the customers to manage their resources. We showed that these databases are poorly protected - the adversaries can take over the accounts for managing the Internet resources and can manipulate the databases, e.g., creating new or removing existing objects - stealthily and causing immediate changes to the customers' resources. %

For our attacks we used different DNS cache poisoning methodologies and compared their applicability and effectiveness for taking over accounts. %
Our work shows that while challenging, our attacks are practical and can be applied against infrastructure of a large fraction of the resource providers to hijack accounts. Our results demonstrate feasibility even with weak off-path adversaries. Certainly, accounts associated with Internet resources are an attractive target also for stronger Man-in-the-Middle adversaries, such as cyber-criminal groups or nation state attackers. %

We described countermeasures for mitigating our off-path attacks for taking over the accounts of customers. Addressing the fundamental problem - easy manipulation of the Internet resources - requires creating policies and revising the Internet management infrastructure as well as techniques for securing the transactions applied over Internet resources. %

\section*{Acknowledgements}
This work has been co-funded by the German Federal Ministry of Education and Research and the Hessen State Ministry for Higher Education, Research and Arts within their joint support of the National Research Center for Applied Cybersecurity ATHENE and by the Deutsche Forschungsgemeinschaft (DFG, German Research Foundation) SFB~1119.

\balance
{\footnotesize
\bibliographystyle{plain}
\bibliography{NetSec,bib,bib2,sec,rfc2,bib3}
}

\end{document}